\documentclass[pdflatex,sn-mathphys-num]{sn-jnl}%

\RequirePackage{xspace}
\makeatletter
\DeclareRobustCommand\onedot{\futurelet\@let@token\@onedot}
\def\@onedot{\ifx\@let@token.\else.\null\fi\xspace}

\def\eg{\emph{e.g}\onedot}

\def\ie{\emph{i.e}\onedot}

\def\vs{\emph{vs}\onedot}

\makeatother

\usepackage{graphicx}%
\usepackage{multirow}%
\usepackage{amsmath,amssymb,amsfonts}%
\usepackage{amsthm}%
\usepackage{mathrsfs}%
\usepackage[title]{appendix}%
\usepackage{xcolor}%
\usepackage{textcomp}%
\usepackage{manyfoot}%
\usepackage{booktabs}%
\usepackage{algorithm}%
\usepackage{algorithmicx}%
\usepackage{algpseudocode}%
\usepackage{listings}%
\usepackage{makecell}
\usepackage{bm}
\usepackage{lineno}
\usepackage{amsmath}
\usepackage{geometry}
\usepackage{graphicx}

\usepackage{booktabs}
\usepackage{multirow}
\usepackage{makecell}
\usepackage{graphicx}
\usepackage{amsmath}
\usepackage{amssymb}
\usepackage{hhline}
\usepackage{algpseudocode}

\theoremstyle{thmstyleone}%

\theoremstyle{thmstyletwo}%

\theoremstyle{thmstylethree}%

\let\oldequation\equation
\let\oldendequation\endequation
\renewenvironment{equation}
  {\linenomathNonumbers\oldequation}
  {\oldendequation\endlinenomath}

\begin{document}

\title[Article Title]{Generative AI models for exploring diffusion dynamics from multi-frame particle tracking}

\author[1]{\fnm{Jianan} \sur{Wei}}\email{weijianan.gm@gmail.com}

\author[2]{\fnm{Qiang} \sur{Zhang}}\email{qiang.zhang.cs@zju.edu.cn}

\author[3]{\fnm{Hao} \sur{Sun}}\email{haosun@ruc.edu.cn}

\author[1]{\fnm{Yi} \sur{Yang}}\email{yangyics@zju.edu.cn}

\author*[1]{\fnm{Wenguan} \sur{Wang}}\email{wenguanwang.ai@gmail.com}

\affil[1]{\orgdiv{State Key Lab of Brain-Machine Intelligence}, \orgname{Zhejiang University}, \orgaddress{\city{Hangzhou}, \state{Zhejiang}, \country{China}}}

\affil[2]{\orgdiv{ZJU-UIUC Institute, International Campus}, \orgname{Zhejiang University}, \orgaddress{\city{Hangzhou}, \state{Zhejiang}, \country{China}}}

\affil[3]{\orgdiv{Gaoling School of Artificial Intelligence}, \orgname{Renmin University of China}, \orgaddress{\city{Beijing}, \country{China}}}

\abstract{
Particle-based volumetric flow measurement techniques have proven their efficacy in quantitatively assessing the dynamic characteristics of unsteady flows within the realm of fluid mechanics.
Despite recent advances in multi-frame particle tracking, an open challenge in this area is maintaining trajectories in temporally under-sampled 3D sequences, where inter-frame displacements are large.
In this work, we develop DiffTR-Flow3D, a diffusion-based model that estimates flow motion from spatiotemporal three-dimensional particle recordings via a stochastic diffusion process.
DiffTR-Flow3D conditions the diffusion process with both global and local particle dynamics and enhances flow estimation by leveraging historical particle trajectories.
Extensive experiments, including numerical simulations and real-world measurements, demonstrate that DiffTR-Flow3D achieves state-of-the-art displacement estimation accuracy compared with learning-based flow estimators, and, when integrated into particle tracking pipelines, substantially improves their long-term tracking performance, while exhibiting robustness and generalization across diverse flow patterns.
This work is expected to provide fluid researchers with an effective tool for the consistent and precise analysis of particle trajectories in temporally under-sampled 3D sequences with large inter-frame displacements.
}
\maketitle

\section{Introduction}\label{sec1}
Tracer particles are widely used in experimental fluid mechanics to measure fluid velocities and derive related quantities, such as accelerations and velocity gradients \cite{schroder20233d}.
Particle Image Velocimetry (PIV) \cite{adrian2005twenty,westerweel2013particle,raffel2018particle} and Particle Tracking Velocimetry (PTV) \cite{malik1993particle} are considered the dominant measurement techniques based on tracer-particle imaging.
PIV typically retrieves the velocity in an Eulerian fashion by tracking groups of particles, while PTV follows a Lagrangian formalism, where each particle is tracked individually, and thus belongs to the broader category of Lagrangian particle tracking (LPT) \cite{schroder20233d}.
Among the techniques built on tracer particles, three-dimensional particle tracking velocimetry (3D PTV) is a volumetric measurement technique that can track a large number of particles over extended timeframes, even under highly turbulent conditions \cite{schroder20233d}.
In a typical multi-camera 3D PTV setup, the iterative particle reconstruction scheme \cite{wieneke2012iterative,Jahn2021} is employed to reconstruct point clouds of 3D particle positions.
Subsequently, a tracking approach is applied to find identical particles at each measured timestep, enabling the construction of 3D particle trajectories.
It provides investigators with data on the flow of interest to theorize and verify flow behavior, leading to applications in several scientific and engineering areas; for instance, investigating interactions between zooplankton swimming and turbulent flow \cite{michalec2017zooplankton}, and resolving transient multiphase flows in porous media \cite{bultreys20244d}.

However, due to tracking ambiguities in high-density fields, conventional frame-to-frame 3D PTV methods often struggle to maintain multi-frame trajectory continuity.
Recently, the Shake-The-Box (STB) method \cite{schanz2016shake,Jahn2021} has advanced 3D PTV by enabling the reconstruction of long and accurate particle trajectories in densely seeded flows, with an almost complete suppression of ghost particles.
However, STB requires time-resolved recordings to implement temporal exploration, which, in conjunction with ``shaking'' operations, facilitates particle position prediction for the reconstruction of 3D trajectories.
As an alternative, Two-pulse STB \cite{novara2023two} extends STB to high-speed scenarios where time-resolved data are unavailable under resource-constrained conditions.
However, when dense Eulerian flow fields are required, the resulting Lagrangian particle data require additional processing, for example using FlowFit \cite{gesemann2016noisy}, which introduces substantial computational overhead \cite{novara2023two}.

3D PTV primarily integrates particle localization and particle linking.
For instance, traditional particle tracking algorithms for large displacements \cite{patel2018rapid,yang2022serialtrack} typically resolve large motion fields by first establishing initial localization and subsequently performing iterative particle linking.
In contrast, STB \cite{schanz2016shake,Jahn2021} integrates particle localization and linking through an iterative process with continuous, intense cross-talk between the two.
Particle localization tries to extract coordinates of individual particles from imaging frames, which is closely tied to specific recording methods.
For example, traditional multi-camera PTV typically processes images via filtering, thresholding, Gaussian fitting for planar positions, and triangulation with calibrated models, which are not applicable to other PTV recording methods like holographic PTV, synthetic aperture PTV, or tomographic PTV \cite{mallery2020dense}.
In contrast, particle linking aims to construct trajectories from 3D particle positions, which is algorithmic and agnostic to hardware and recording factors.
In this paper, we decouple these components to focus on spatiotemporal motion modeling for multi-frame particle tracking.
Using established point clouds, we aim to estimate particle displacements and predict particle trajectories from temporally under-sampled multi-frame recordings through a diffusion-inspired denoising formulation.

Traditional iterative-based particle tracking algorithms \cite{patel2018rapid,yang2022serialtrack} are time-consuming and require extensive manual parameter configuration.
These limitations have driven growing interest in deep learning methods for estimating flow motion in particle tracking.
Neural networks have been utilized in planar PTV prior to the 21st century \cite{knaak1997hopfield,hassan1997new,labonte1999new}, serving as robust tools for the particle linking process \cite{ohmi2006particle}.
Building upon these foundations, the PIV community has integrated convolutional neural networks (CNNs) into the estimation of flow motion \cite{rabault2017performing,lee2017piv,cai2019dense}, subsequently shifting towards optical flow methods as the prevailing approach \cite{cai2019deep,cai2019particle,lagemann2021deep}.
Paralleling this evolution, recent 3D PTV research has increasingly focused on 3D optical flow estimation frameworks tailored for point cloud data \cite{liang2021deepptv,liang2023recurrent}.
Despite their strong performance, these learning-based methods are designed for double-frame settings and can suffer from error accumulation and performance degradation when applied sequentially to multi-frame tracking.
Moreover, they reduce the intricate task of flow trajectory estimation to a matching problem between two sets of points, thereby overlooking the continuous displacement process of particles between frames.
In practice, most natural flows are unsteady, turbulent, and three-dimensional, which create intricate and diverse trajectory patterns.
To address this complexity, we draw inspiration from diffusion-based models \cite{ho2020denoising} and treat the estimation of flow trajectories as a diffusion-based stochastic process.

Motivated by these observations, we focus on multi-frame particle tracking in scenarios where temporal under-sampling leads to large inter-frame displacements and propose DiffTR-Flow3D, a diffusion-based model that estimates 3D flow by predicting displacement fields while using stochastic noise modeling to regularize and guide the estimation process.
Specifically, DiffTR-Flow3D consists of three key components:
(1) a robust two-step flow estimator designed to handle diverse displacement distributions in particle dynamics;
(2) a conditional diffusion-based denoising strategy that replaces conventional recurrent updates with a noise-conditioned flow refinement process;
(3) a temporal fitting module implemented with a regularized polynomial function to exploit historical particle trajectories, thereby guiding the reverse denoising process with temporal information.

Since we exclusively focus on displacement estimation for linking multiple temporal instances of identical particles, our model is trained using a high-quality synthetic 3D point cloud dataset obtained from computational fluid dynamics (CFD) simulations.
Our synthetic generation protocol covers diverse flow patterns and includes a controlled level of particle corruption, enabling the supervised learning of DiffTR-Flow3D with highly accurate ground-truth flow.
Extensive experiments are carried out to rigorously validate our solution.
Specifically, we evaluate our DiffTR-Flow3D as a trajectory estimator on the FluidFlow3D dataset \cite{liang2023recurrent} and our newly created TR-Flow3D dataset under both double-frame and multi-frame settings.
Furthermore, we demonstrate the method's robustness against elevated particle loss levels in the test data and provide a detailed assessment at each timestep on the TR-Flow3D dataset.
Additionally, by integrating the pre-trained DiffTR-Flow3D with traditional 3D PTV pipelines, we assess its generalization and robustness for particle tracking on the CylinderFlow dataset and the experimental test case of hydrogel shear.
The superior performance of DiffTR-Flow3D across all these experiments highlights its capability and adaptability to support particle association through accurate motion prediction across diverse flow conditions.

\begin{figure}[!htbp]
 	\begin{center}
 	\includegraphics[width=1.0\columnwidth]{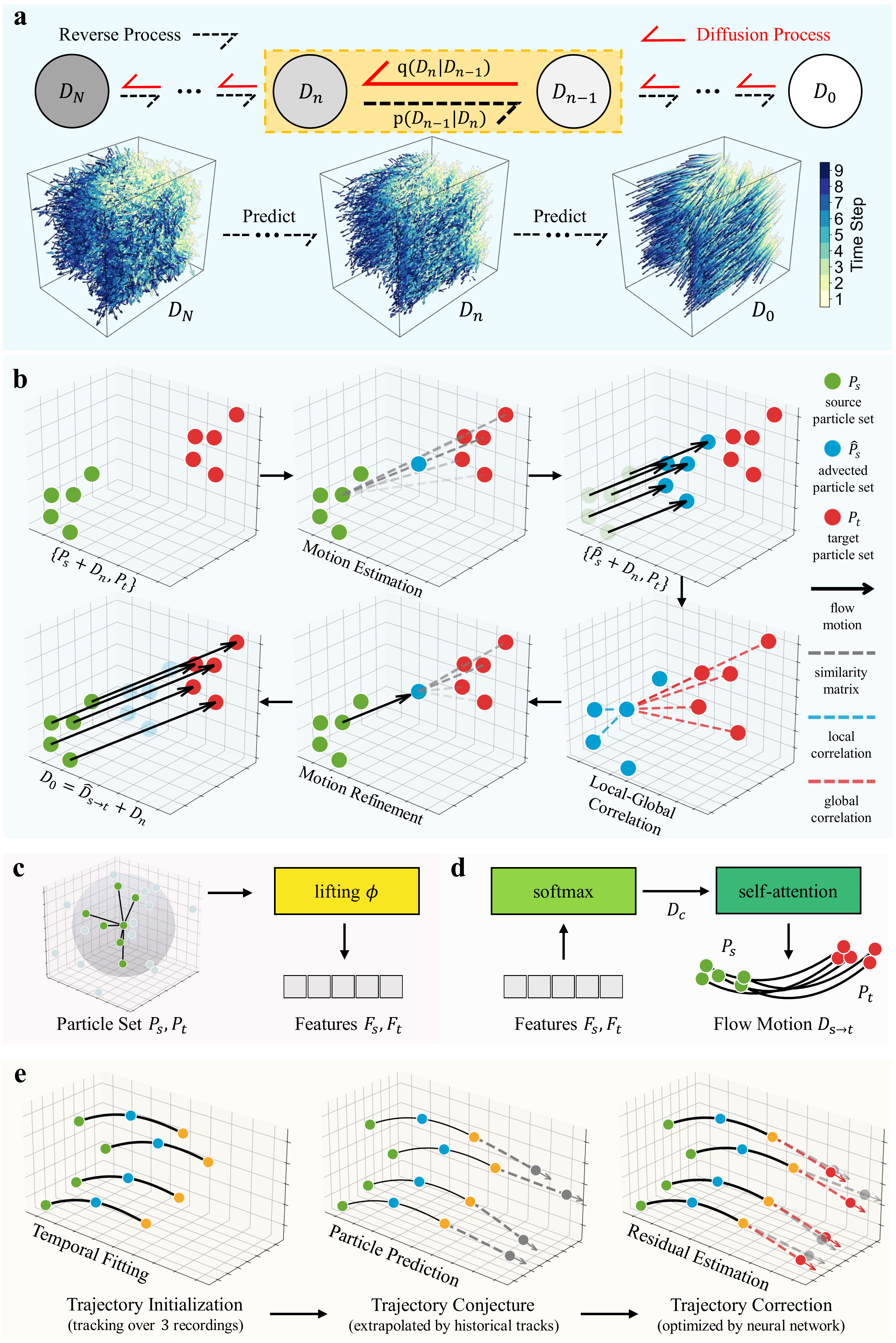}
 	\end{center}
 	\caption{\textbf{Overview of DiffTR-Flow3D.}
	\textbf{a,} Overview of the noise perturbation and reverse denoising processes in DiffTR-Flow3D. The forward diffusion process, $q(d_{1:N} | d_{0})$, progressively perturbs the flow motion structure $d_0 \sim q(d)$ over $N$ steps to approximate a Gaussian noise distribution. Conversely, the reverse process, $p_{\theta}(d_{n-1} | d_{n})$, iteratively eliminates Gaussian noise from the corrupted flow $d_N$, thereby guiding particles toward their target destinations.
	\textbf{b,} The flow motion estimation pipeline in DiffTR-Flow3D. This module employs a two-stage flow motion estimator designed to handle diverse displacement distributions in particle dynamics. It first performs a coarse flow estimation between two point sets, followed by a refinement stage that eliminates position offsets using soft alignment operations to capture both local and global correlations.
	\textbf{c,} Illustration of the feature extraction pipeline in DiffTR-Flow3D.
	\textbf{d,} The workflow of motion estimation/refinement employed by DiffTR-Flow3D.
	\textbf{e,} Schematic of the multi-frame trajectory prediction scheme in DiffTR-Flow3D. This scheme first establishes particle trajectories via flow estimation over three consecutive timesteps. Subsequently, particle positions for the next timestep are predicted using temporal fitting functions derived from prior trajectory information. Finally, the proposed flow estimator calculates the residual motion, which is used to correct the trajectories.
}
 	\label{fig:method}
 \end{figure}

\section{DiffTR-Flow3D}\label{sec2}

\noindent\textbf{Problem setup.}
Given the reference frame initialized by the first frame, our method predicts particle trajectories in a cumulative mode via an alternating two-step process:
(1) the subsequent frame is aligned with the reference frame to directly compute cumulative displacement fields through the proposed DiffTR-Flow3D;
(2) the reference frame is updated by integrating the estimated displacement fields.
The resulting position estimates form predicted particle trajectories, which guide particle association and trajectory linking in the integrated tracking pipelines described in Section ~\ref{res:generalize}.
Let $P_s \in \mathbb{R}^{N_s \times 3}$ and $\mathcal{P}_t=\{P_t^{(k)} \in \mathbb{R}^{N_t \times 3}\}^{K}_{k=1}$ denote the source point set and the target point sets, where $N_s$ and $N_t$ are the numbers of points in the source and target point sets, respectively, and $K+1$ is the total number of frames.
Our purpose is to develop a neural network that estimates multi-frame particle displacements $\mathcal{D}_{s \rightarrow t} = \{D_{s \rightarrow t}^{(k)} \in \mathbb{R}^{N_s \times 3 }\}^{K}_{k=1}$ by mapping source points to their corresponding positions in each target frame over time.
This process is enhanced by a conditional diffusion paradigm (Fig. \ref{fig:method}a), where the flow motion structure $d_0 \sim q(d)$ is progressively perturbed over $N$ steps to approximate a Gaussian noise distribution, \ie, the diffusion process $q(d_{1:N} | d_{0})$, while a reverse process  $p_{\theta}(d_{n-1} | d_{n})$  is learned to restore motion patterns by eliminating the noise.

\noindent\textbf{Flow motion estimation.}
Particles in particle tracking commonly exhibit significant velocity fluctuations, leading to displacements across various scales.
This complexity poses a substantial challenge in achieving precise flow estimation in a single step.
Following previous recurrent methods \cite{wei2021pv,liang2023recurrent} for particle tracking, our approach also employs a coarse-to-fine structure.
As demonstrated in Fig. \ref{fig:method}b, DiffTR-Flow3D utilizes a powerful two-stage flow motion estimator to handle diverse displacement distributions in particle dynamics:
first, it coarsely estimates flow motion between two point sets;
then, it refines the flow motion by eliminating the position offsets with soft alignment operations that capture local and global correlations.

\noindent\textbf{Multi-frame trajectory prediction.}
Due to the disordered nature of point clouds, 3D PTV typically performs particle linking by sequentially correlating particle positions between two consecutive frames.
For scenarios involving large displacements and sparse particle distributions, double-frame tracking methods  (\eg, T-PT \cite{patel2018rapid} and SerialTrack \cite{yang2022serialtrack}) remain prevalent, due to the limited utility of temporal information.
Under high-temporal-sampling scenarios, the small inter-frame displacement enables multi-frame algorithms such as STB \cite{schanz2016shake,Jahn2021} to leverage temporal fitting to initialize particle positions in the next timestep for enhanced particle tracking performance.
In light of this, our approach predicts multi-frame particle trajectories via a three-stage paradigm (Fig. \ref{fig:method}e):
(1) we first perform flow estimation to initialize trajectories over the first few consecutive frames;
(2) the derived trajectory information is then leveraged to predict coarse particle positions for the subsequent timestep via temporal fitting;
(3) these coarse predictions serve as initial conditions for DiffTR-Flow3D, thereby simplifying large displacement estimation into a residual offset estimation.
Notably, the integration of diffusion processes enables DiffTR-Flow3D to handle noisy temporal predictions, as it employs Gaussian noise to perturb flow motion during training.

\section{Results}\label{sec3}
Following the evaluation protocol established in \cite{lagemann2021deep,liang2023recurrent}, we evaluate our method on two benchmarks under double-frame and multi-frame settings, with state-of-the-art scene flow estimation approaches (\eg, FLOT \cite{puy2020flot}, PointPWC-Net \cite{wu2020pointpwc}, PV-RAFT \cite{wei2021pv}, 3DFlow \cite{wang2022matters}, GMSF \cite{zhang2024gmsf}, DiffSF \cite{zhang2024diffsf}), and an advanced 3D flow motion learner (\ie, GotFlow3D \cite{liang2023recurrent}) to validate the architectural superiority of DiffTR-Flow3D as a deep learning-based flow estimator for 3D PTV.
All experiments are conducted on the same hardware platform (two NVIDIA RTX 4090 GPUs) to ensure a fair comparison. It is worth noting that, except for evaluations conducted on the FluidFlow3D dataset, our model is trained only once on the TR-Flow3D training dataset. We utilize the model snapshot from the final training iteration for evaluations on the TR-Flow3D test set, the CylinderFlow dataset, and the experimental data, without any case-specific fine-tuning or additional training. For the FluidFlow3D benchmark, we strictly follow the protocol of \cite{liang2023recurrent} by retraining the models on their provided training set.
We employ EPE, ACC Strict, ACC Relax, and Outliers to evaluate the accuracy of the predicted particle displacements, and, for the integrated PTV pipelines, the tracking accuracy \(E_y\) to evaluate whether particle identities are correctly maintained over multiple frames. All metrics are defined in Section~\ref{sec:eval}.
First, we evaluate the performance of DiffTR-Flow3D under the conventional double-frame setting \cite{liang2021deepptv,liang2023recurrent}, which simulates the initial phase of multi-frame particle tracking without trajectory information on the FluidFlow3D and TR-Flow3D datasets.
Next, we assess DiffTR-Flow3D's capability under a more challenging and realistic multi-frame setting, which requires maintaining precise and swift displacement estimation with prior trajectories on the TR-Flow3D dataset.
Additionally, we test the model's robustness under challenging conditions, \ie, particle loss, and large displacements due to extended inter-frame time intervals.
Moreover, to evaluate DiffTR-Flow3D’s generalization capability and its contribution to long-term tracking when integrated into established particle tracking pipelines (\ie, T-PT \cite{patel2018rapid} and SerialTrack \cite{yang2022serialtrack}), we compare these pipelines with and without DiffTR-Flow3D on a public numerical dataset (referred to as CylinderFlow) \cite{KHOJASTEH2022107725}  as well as a real experimental test case (hydrogel shear) \cite{yang2022serialtrack}.
Specifically, T-PT and SerialTrack, originally designed for double-frame particle linking, are extended to multi-frame tracking through post-processing trajectory stitching.

\begin{table}[t]
\caption{\textbf{Comparison of particle motion estimation performance on the FluidFlow3D dataset under the double-frame setting.}}\label{tab1}%
\begin{tabular}{@{}llllll@{}}
\toprule
Methods & Params& EPE$\downarrow$& ACC Strict$\uparrow$& ACC Relax$\uparrow$& Outliers$\downarrow$\\
\midrule
GMSF\cite{zhang2024gmsf}& 7.07M& 0.16313& 20.40\%& 27.73\%& 72.27\% \\
FLOT \cite{puy2020flot}& 0.11M& 0.05865& 24.99\%& 45.59\%& 54.41\% \\
PointPWC-Net \cite{wu2020pointpwc}& 7.72M& 0.01723& 46.75\%& 71.61\%& 28.39\% \\
PV-RAFT \cite{wei2021pv}& 0.19M& 0.01651& 72.98\%& 83.69\%& 16.31\% \\
3DFlow\cite{wang2022matters}& 4.77M& 0.02699& 84.91\%& 88.29\%& 11.71\% \\
DiffSF\cite{zhang2024diffsf}& 9.28M&  0.01403 & 95.52\% & 96.24\% & 3.76\% \\
GotFlow3D \cite{liang2023recurrent}& 1.44M& 0.00487& 93.15\%& 96.38\%& 3.62\% \\
\midrule
DiffTR-Flow3D& 9.28M& \textbf{0.00230}& \textbf{99.25\%}& \textbf{99.36\%}& \textbf{0.64\%}  \\
\botrule
\end{tabular}
\end{table}

\begin{table}[t]
\caption{\textbf{Comparison of particle motion estimation performance}  on the TR-Flow3D dataset under the double-frame setting.}\label{tab2}%
\begin{tabular}{@{}llllll@{}}
\toprule
Methods & Params& EPE$\downarrow$& ACC Strict$\uparrow$& ACC Relax$\uparrow$& Outliers$\downarrow$\\
\midrule
GMSF \cite{zhang2024gmsf}& 7.07M& 0.16993& 24.78\%& 33.15\%& 66.85\% \\
FLOT \cite{puy2020flot}& 0.11M& 0.05291& 41.43\%& 57.15\%& 42.85\% \\
PointPWC-Net \cite{wu2020pointpwc}& 7.72M& 0.05228& 44.38\%& 62.06\%& 37.94\% \\
PV-RAFT \cite{wei2021pv}& 0.19M& 0.05220& 32.25\%& 51.51\%& 48.49\% \\
3DFlow \cite{wang2022matters}& 4.77M& 0.03441& 74.65\%& 82.31\%& 17.69\% \\
DiffSF\cite{zhang2024diffsf}& 9.28M& 0.01526 & 94.90\% & 95.98\% & 4.02\% \\
GotFlow3D \cite{liang2023recurrent}& 1.44M& 0.01153& 88.19\%& 94.02\%& 5.98\% \\
\midrule
DiffTR-Flow3D& 9.28M& \textbf{0.00275}& \textbf{98.47\%}& \textbf{98.85\%}& \textbf{1.15\%} \\
\botrule
\end{tabular}
\end{table}

\par\vspace{12pt}
\subsection{Double-frame motion estimation for particle tracking}
As the foundation of 3D PTV, double-frame particle tracking provides essential data for continuous trajectory reconstruction.
Thus, we discuss the performance of DiffTR-Flow3D against its competitors in accurately estimating flow between two adjacent frames, \ie, under the double-frame setting.
First, we report the results on the FluidFlow3D dataset \cite{liang2023recurrent}, which contains only pairs of particle frames to be tracked.
As shown in Table \ref{tab1}, DiffTR-Flow3D outperforms other state-of-the-art scene flow estimation methods \cite{puy2020flot,wu2020pointpwc,wei2021pv,wang2022matters,liang2023recurrent,zhang2024gmsf,zhang2024diffsf} by a large margin across all settings.
Particularly, DiffTR-Flow3D achieves scores exceeding 99\% in both ACC Strict and ACC Relax metrics, outperforming GotFlow3D by 6.10 percentage points in ACC Strict.
To explore the performance ceiling of the proposed DiffTR-Flow3D, we also conduct experiments on our newly created TR-Flow3D dataset.
This dataset is more challenging due to its extended and more varied time intervals, leading to larger and more heterogeneous particle displacements.
Most baseline methods, as evidenced in Table \ref{tab2}, exhibit a decline in performance; conversely, our method sustains its performance at 98.47\% for ACC Strict and 98.85\% for ACC Relax, which demonstrates its superior robustness and reliability.

\par\vspace{12pt}
\subsection{Multi-frame motion estimation for particle tracking}
Then, we assess DiffTR-Flow3D against its competitors on the TR-Flow3D dataset in a more challenging multi-frame scenario, which necessitates predicting particle positions over successive frames from the initial frame to the final frame.
As shown in Table \ref{tab3}, DiffTR-Flow3D also outperforms other methods by a significant margin, reaching 98.48\% in ACC Strict and 98.72\% in ACC Relax.
In particular, in terms of End-Point-Error (EPE), it reduces the error of the strongest baseline, GotFlow3D, by more than sixfold (0.00316 vs. 0.02015), which suggests that DiffTR-Flow3D can maintain accurate particle motion estimates over multiple frames.
We attribute this performance gap to the limitations of recent optical flow estimators such as PV-RAFT~\cite{wei2021pv} and GotFlow3D \cite{liang2023recurrent}.
Although these methods can be adapted for iterative tracking, they rely on sequential propagation without temporal constraints, making them susceptible to cumulative drift.
In contrast, DiffTR-Flow3D addresses this limitation by leveraging historical trajectories, which provide a robust initial guess for particle positions in the next timestep.
Furthermore, we investigate the performance of the models across different flow scenarios.
Table \ref{tab4} demonstrates that our method achieves robust results across all six flow scenarios.
A comparative analysis with other established methods (see Table \ref{tab:comparison_pattern} for more results) confirms the superior performance of our approach in these diverse conditions.

\begin{figure}[t]
 	\begin{center}
 	\includegraphics[width=1.0\columnwidth]{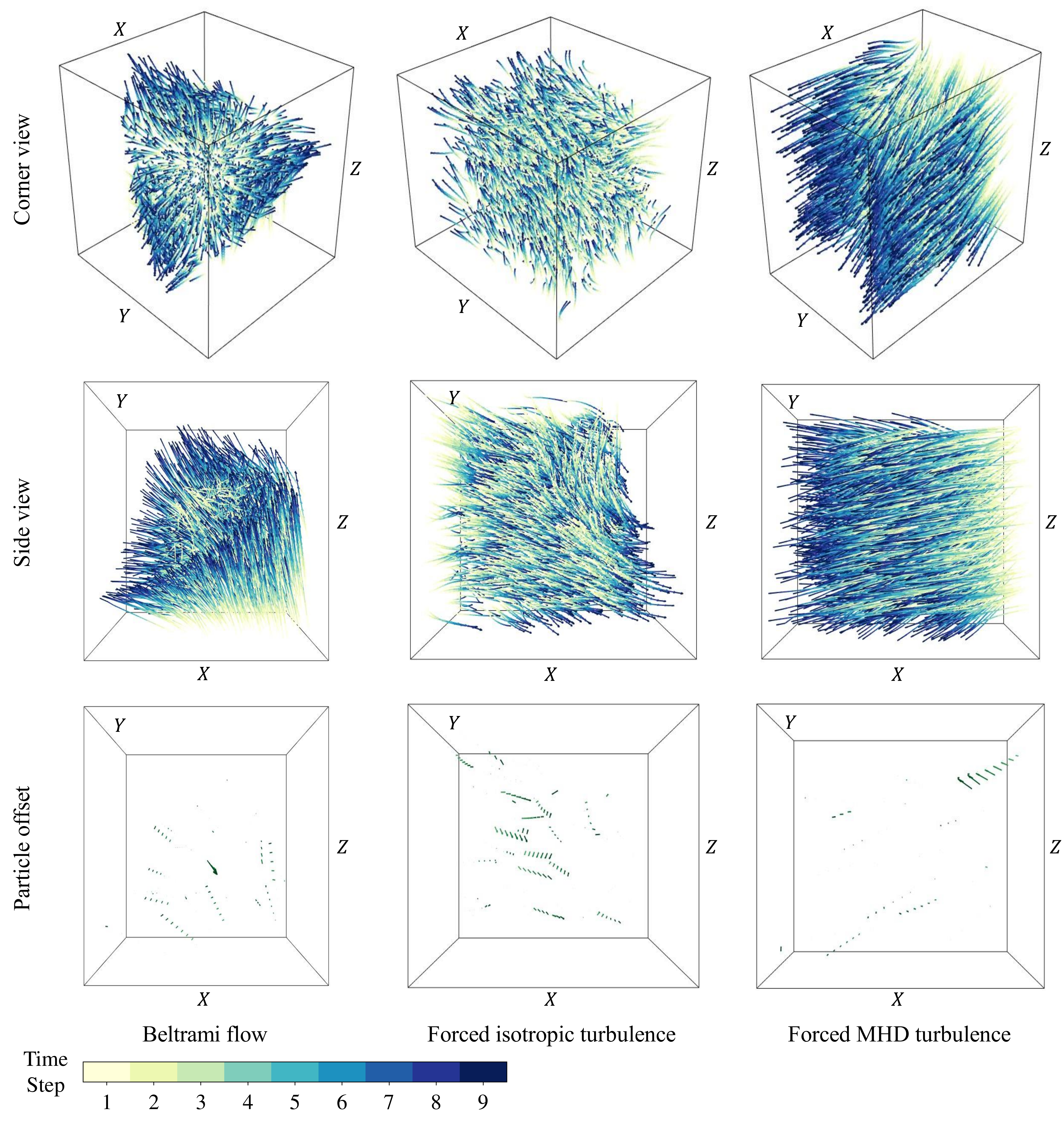}
 	\end{center}
	\caption{\textbf{Predicted particle trajectories for three test cases on the TR-Flow3D dataset.}}
 	\label{fig:visualization}
\end{figure}

Fig. \ref{fig:visualization} illustrates three representative examples, including Beltrami flow, forced isotropic flow, and forced MHD turbulence.
It can be seen that the flow motion estimated by DiffTR-Flow3D exhibits distinct flow characteristics, such as vortex structures and local isotropy.
In addition, we provide the position offsets computed relative to the ground truth, displaying only trajectories for which the EPE exceeds 0.01.
Notably, only a few trajectories exhibit significant errors, and these errors remain stable with minor changes across subsequent frames.
We attribute this stability to the utilization of prior trajectory information.

\begin{table}[t]
\caption{\textbf{Comparison of particle motion estimation performance  on the TR-Flow3D dataset under the multi-frame setting.}}\label{tab3}%
\begin{tabular}{@{}llllll@{}}
\toprule
Methods & Params& EPE$\downarrow$& ACC Strict$\uparrow$& ACC Relax$\uparrow$& Outliers$\downarrow$\\
\midrule
GMSF\cite{zhang2024gmsf}& 7.07M&  0.18789& 25.11\%& 33.03\%& 66.97\% \\
FLOT\cite{puy2020flot}& 0.11M& 0.06051& 41.03\%& 54.54\%& 45.46\% \\
PointPWC-Net\cite{wu2020pointpwc}& 7.72M& 0.07101& 50.14\%& 63.61\%& 36.39\% \\
PV-RAFT\cite{wei2021pv}& 0.19M& 0.06379& 32.09\%& 49.41\%& 50.59\% \\
3DFlow\cite{wang2022matters}& 4.77M& 0.06360& 63.07\%& 75.52\%& 24.48\% \\
GotFlow3D\cite{liang2023recurrent}& 1.44M& 0.02015& 85.26\%& 90.31\%& 9.69\% \\
\midrule
DiffTR-Flow3D& 9.28M& \textbf{0.00316}& \textbf{98.48\%}& \textbf{98.72\%}& \textbf{1.28\%} \\
\botrule
\end{tabular}
\end{table}

\begin{table}[t]
\caption{\textbf{Particle motion estimation performance of DiffTR-Flow3D on various flow field patterns on the TR-Flow3D testing dataset.}}\label{tab4}%
\begin{tabular}{@{}llllll@{}}
\toprule
Flow field pattern& EPE$\downarrow$& ACC Strict$\uparrow$& ACC Relax$\uparrow$& Outliers$\downarrow$\\
\midrule
Uniform flow& 0.00028& $100.0$\%& $100.0$\%&  $0.00$\% \\
Transitional boundary flow& 0.00078& 99.99\%&  $100.0$\%&  $0.00$\% \\
Turbulent channel flow& 0.00078& 99.91\%& 99.97\%& 0.03\% \\
Beltrami flow& 0.00390& 98.43\%& 98.79\%& 1.21\% \\
Forced MHD turbulence& 0.00148& 97.55\%& 98.66\%& 1.34\% \\
Forced isotropic turbulence& 0.00803& 95.68\%& 96.24\%& 3.76\% \\
\botrule
\end{tabular}
\end{table}

\par\vspace{12pt}
\subsection{Robustness analysis of DiffTR-Flow3D}
To provide a comprehensive analysis of DiffTR-Flow3D's robustness, we evaluate it from two perspectives:
1) \textit{Particle loss ratio}.
In each frame, a randomly selected proportion of particles is replaced with points drawn from a Gaussian distribution within the observation volume to mimic the loss of particle positions induced by systematic errors.
Notably, the particle loss progressively accumulates over successive frames;
2) \textit{Timestep}. We also present a detailed assessment of flow estimation performance at each timestep along the particle trajectories.

\begin{figure}[t]
 	\begin{center}
 	\includegraphics[width=1.0\columnwidth]{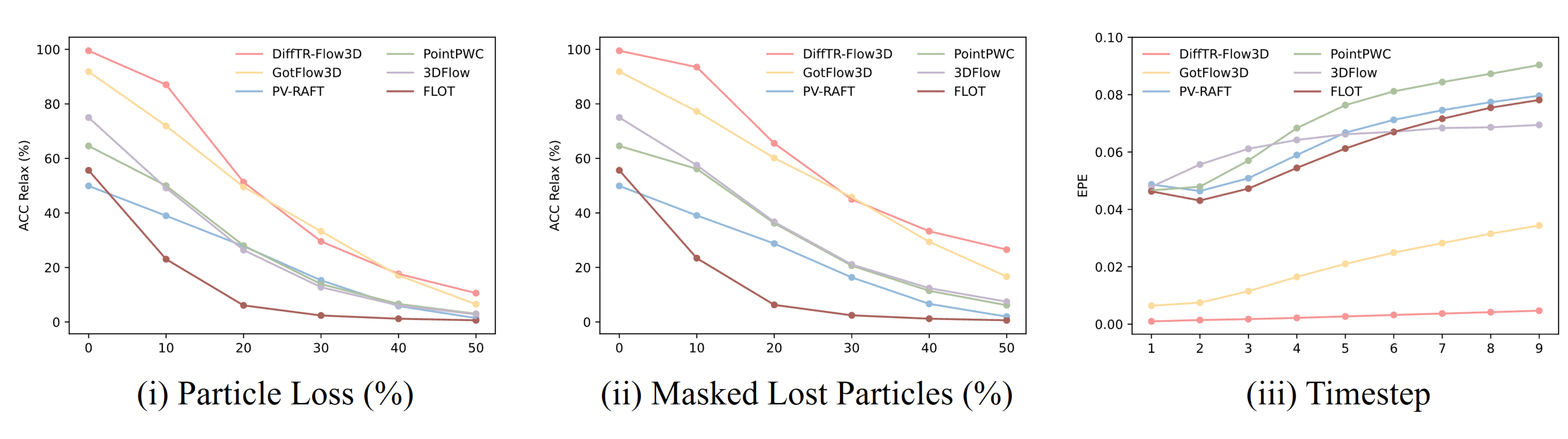}
 	\end{center}
 	\caption{\textbf{Performance of different flow estimators on the TR-Flow3D test dataset.} (i) and (ii) depict the performance under varying particle loss ratios, with (ii) evaluating metrics only on particles that are tracked (for analytical purposes). Moreover, (iii) illustrates the variation in EPE with time step increasing.}
 	\label{fig:condition}
\end{figure}

\noindent\textbf{Particle loss ratio per frame.}
The reconstruction of 3D particles from 2D images faces challenges such as optical distortions, out-of-plane motion, and particle overlaps, which introduce missed detections or falsely reconstructed particles that do not exist in the original particle cloud.
The former leads to fragmented trajectories and disrupted temporal continuity in tracking, while the latter generates erroneous trajectories that propagate and accumulate over time, which is more detrimental to particle tracking.
To simulate these events, we replace a certain proportion of particles with points drawn from a Gaussian distribution within the observation volume specified in Table \ref{tab:dataset}.
As shown in Fig. \ref{fig:condition}i, it can be observed that the ACC Relax of all methods declines with increasing particle loss ratios; however, when the loss ratio reaches 50\% (approaching the upper limit typically observed in real experiments), DiffTR-Flow3D still exhibits the best performance.
Notably, when we focus solely on trackable particles (excluding lost particles), the performance advantage of DiffTR-Flow3D is markedly more pronounced under conditions with higher particle loss ratios (as shown in Fig. \ref{fig:condition}ii), thereby underscoring its superior robustness.

\noindent\textbf{Timestep along particle trajectories.}
To illustrate the robustness of different flow estimators against cumulative error during multi-frame trajectory prediction, Fig. \ref{fig:condition}iii delineates the EPE at each frame throughout the evaluated sequence on the TR-Flow3D dataset.
It can be observed that all competitors struggle with multi-frame scenarios: while these methods maintain acceptable performance during the initial three frames, a pronounced decline in performance becomes apparent thereafter.
For instance, GotFlow3D exhibits favorable performance initially; however, by the end, the EPE increases to more than five times its initial value.
In contrast, our DiffTR-Flow3D sustains stable performance across the entire trajectory, achieving results at the end that are even superior to GotFlow3D's performance at the beginning.

\begin{table}[t]
\caption{\textbf{Tracking accuracy $E_y$ (detailed in Eq. \ref{eq:tracking_acc}) for the CylinderFlow dataset.}}\label{tab:cylinderflow}%
\begin{tabular}{@{}l|lll|lll|lll@{}}
\toprule %
& \multicolumn{3}{c|}{$\Delta t=30$}& \multicolumn{3}{c|}{$\Delta t=60$}& \multicolumn{3}{c}{$\Delta t=90$}\\
\multirow{-2}{*}{$n_p=3,940$}& \texttt{First}& \texttt{Last}& \texttt{Mean}& \texttt{First}& \texttt{Last}& \texttt{Mean}& \texttt{First}& \texttt{Last}& \texttt{Mean}\\
\midrule
SerialTrack& 98.77& 70.69& 82.84& 94.80& 61.35& 75.18& 90.17& 52.95& 67.67 \\
$\text{SerialTrack}^{\dag}$& 99.56& 90.89& 94.59& 98.38& 89.99& 93.42& 96.89& \textbf{87.28}& 91.52 \\
T-PT& 99.72& 95.38& 97.43& 95.67& 79.49& 86.62& 82.54& 58.00& 67.67 \\
$\text{T-PT}^{\dag}$& \textbf{100.00}& \textbf{98.16}& \textbf{99.26}& \textbf{99.92}& \textbf{96.85}& \textbf{98.63}& \textbf{99.58}& 82.20& \textbf{93.40} \\
\midrule

& \multicolumn{3}{c|}{$\Delta t=30$}& \multicolumn{3}{c|}{$\Delta t=60$}& \multicolumn{3}{c}{$\Delta t=90$}\\
\multirow{-2}{*}{$n_p=9,850$}& \texttt{First}& \texttt{Last}& \texttt{Mean}& \texttt{First}& \texttt{Last}& \texttt{Mean}& \texttt{First}& \texttt{Last}& \texttt{Mean}\\
\midrule
SerialTrack& 97.92& 68.52& 81.00& 94.69& 59.02& 73.62& 88.76& 48.93& 65.11 \\
$\text{SerialTrack}^{\dag}$& 99.45& 88.08& 93.80& 98.24& 87.19& 92.60& 95.89& \textbf{83.04}& \textbf{89.41} \\
T-PT& 99.65& 93.50& 96.48& 91.18& 70.34& 79.31& 73.36& 50.64& 59.32 \\
$\text{T-PT}^{\dag}$& \textbf{99.99}& \textbf{97.58}& \textbf{99.02}& \textbf{99.76}& \textbf{89.97}& \textbf{96.71}& \textbf{98.34}& 74.22& 87.38 \\
\midrule

& \multicolumn{3}{c|}{$\Delta t=30$}& \multicolumn{3}{c|}{$\Delta t=60$}& \multicolumn{3}{c}{$\Delta t=90$}\\
\multirow{-2}{*}{$n_p=19,700$}& \texttt{First}& \texttt{Last}& \texttt{Mean}& \texttt{First}& \texttt{Last}& \texttt{Mean}& \texttt{First}& \texttt{Last}& \texttt{Mean}\\
\midrule
SerialTrack& 97.97& 69.08& 81.98& 93.99& 59.38& 73.92& 88.15& 50.13& 65.48 \\
$\text{SerialTrack}^{\dag}$& 99.24& 85.75& 92.95& 97.84& \textbf{82.81}& 90.28& \textbf{93.77}& \textbf{73.96}& \textbf{83.58} \\
T-PT& 98.96& 91.86& 95.24& 86.61& 65.74& 74.34& 67.96& 50.08& 56.98 \\
$\text{T-PT}^{\dag}$& \textbf{99.93}& \textbf{96.94}& \textbf{98.74}& \textbf{99.35}& 80.11& \textbf{91.48}& 90.86& 62.78& 76.03 \\
\botrule
\end{tabular}
$\text{SerialTrack}^{\dag}$ and $\text{T-PT}^{\dag}$ represent the integration of DiffTR-Flow3D into SerialTrack and T-PT, respectively.
The best results are highlighted in bold; $\Delta t$ signifies the time interval between frames; $n_p$ indicates the number of particles sampled from the CylinderFlow dataset. \texttt{First} denotes the tracking accuracy of the initial trajectory segment (between the first and second frames); \texttt{Last} measures the association accuracy between the first and final frames; and \texttt{Mean} represents the average tracking accuracy computed across all time steps.
\end{table}

\begin{figure}[t]
 	\begin{center}
 	\includegraphics[width=1.0\columnwidth]{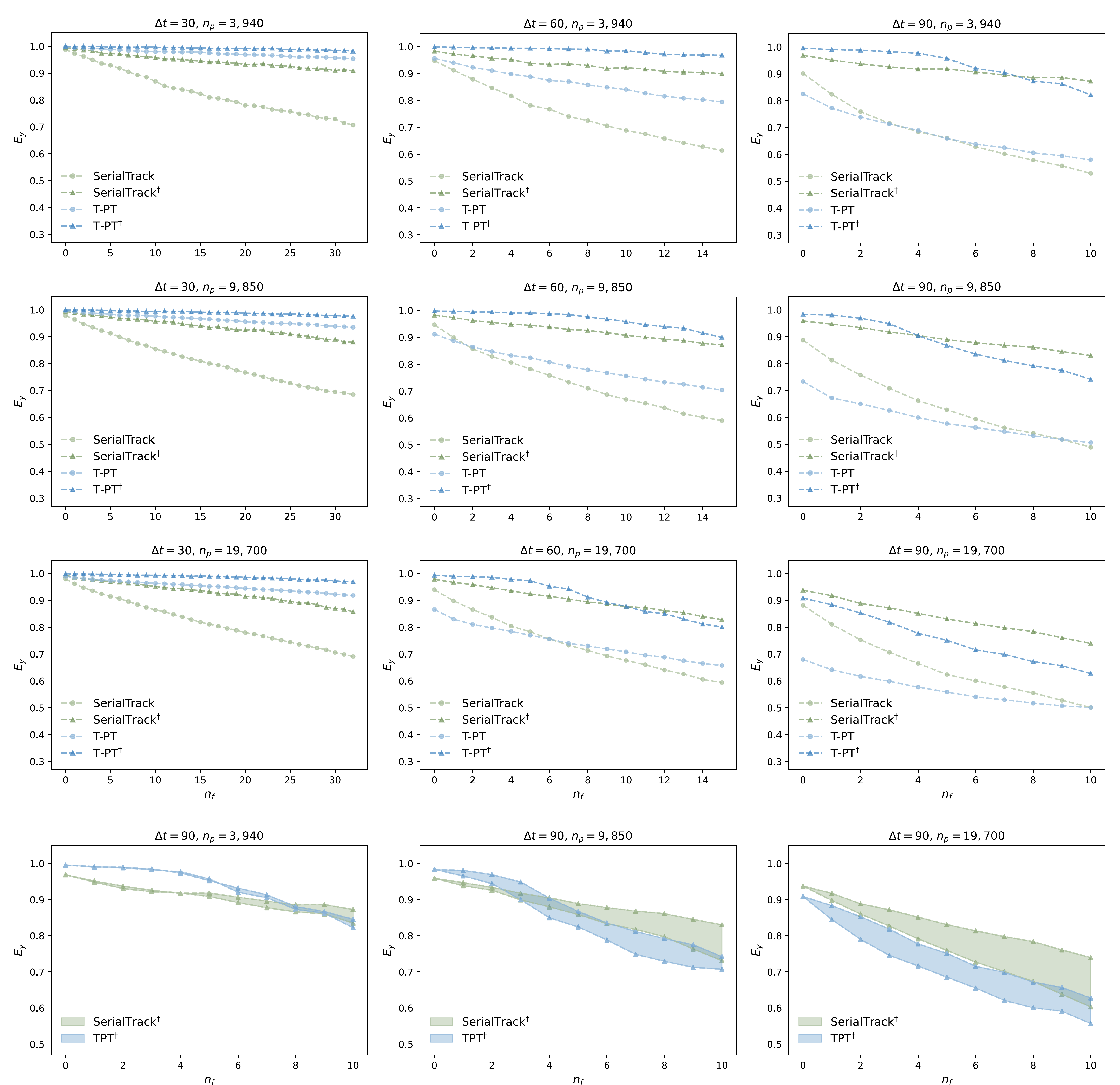}
	\put(-370,350){\scalebox{1.00}{\textbf{a}}}
	\put(-370,80){\scalebox{1.00}{\textbf{b}}}
 	\end{center}
 	\caption{\textbf{Results for the CylinderFlow dataset.} \textbf{a,} Comparison with 3D particle tracking algorithms. $\text{SerialTrack}^{\dag}$ and $\text{T-PT}^{\dag}$ represent the integration of DiffTR-Flow3D into SerialTrack and T-PT, respectively; $n_f$ denotes the timestep; $E_y$ indicates the tracking accuracy (detailed in Eq. \ref{eq:tracking_acc}). \textbf{b,} Impact of temporal information on performance. The upper boundary represents the integration of the complete DiffTR-Flow3D with trajectory information, whereas the lower boundary corresponds to the method excluding trajectory information.}
 	\label{fig:cylinderflow}
\end{figure}

\par\vspace{12pt}
\subsection{Generalization to unseen scenarios} \label{res:generalize}
To assess the robustness and out-of-distribution generalization of DiffTR-Flow3D, we directly evaluate it on the unseen CylinderFlow dataset (high Reynolds numbers) and experimental data from hydrogel shear testing (sparse particle density) without any retraining or fine-tuning.
We first assess the model on the CylinderFlow dataset following the experimental protocol in \cite{liang2023recurrent}.
However, unlike GotFlow3D, each case in our evaluation includes multiple frames, \ie, the numbers of frames $n_f$ for $\Delta t=30$, $\Delta t=60$, and $\Delta t=90$ are $n_f=34$, $n_f=17$, and $n_f=12$, respectively.
We adopt two open-source 3D particle tracking algorithms, T-PT \cite{patel2018rapid} and SerialTrack \cite{yang2022serialtrack}, to continuously track the identified particles from the first frame through all subsequent frames, serving as benchmarks.
Because STB \cite{schanz2016shake,Jahn2021} is primarily designed for time-resolved tracking under relatively small inter-frame displacements, it is not used as the primary baseline in this large-displacement evaluation; its results are reported in Appendix \ref{secA}, Table \ref{tab:app_cylinderflow} for completeness.
Specifically, we first perform particle matching between consecutive frames based on T-PT or SerialTrack, and then link the initial frame to all subsequent ones through a chain of matches.
DiffTR-Flow3D is capable of providing trajectories for observed particles; however, it does not offer a definitive one-to-one matching across particle frames.
Therefore, following the evaluation setup in GotFlow3D \cite{liang2023recurrent}, we utilize DiffTR-Flow3D as a customized motion predictor to initialize existing 3D PTV algorithms with accurate flow fields for T-PT and SerialTrack, thereby formulating $\text{T-PT}^{\dag}$ and $\text{SerialTrack}^{\dag}$.
The matching process in these algorithms is refined by shifting from inter-frame matching to a matching between the estimated trajectories and the observed particle positions, thereby mitigating error accumulation inherent in existing algorithms.

\begin{figure}[!tbp]
 	\begin{center}
 	\includegraphics[width=1.0\columnwidth]{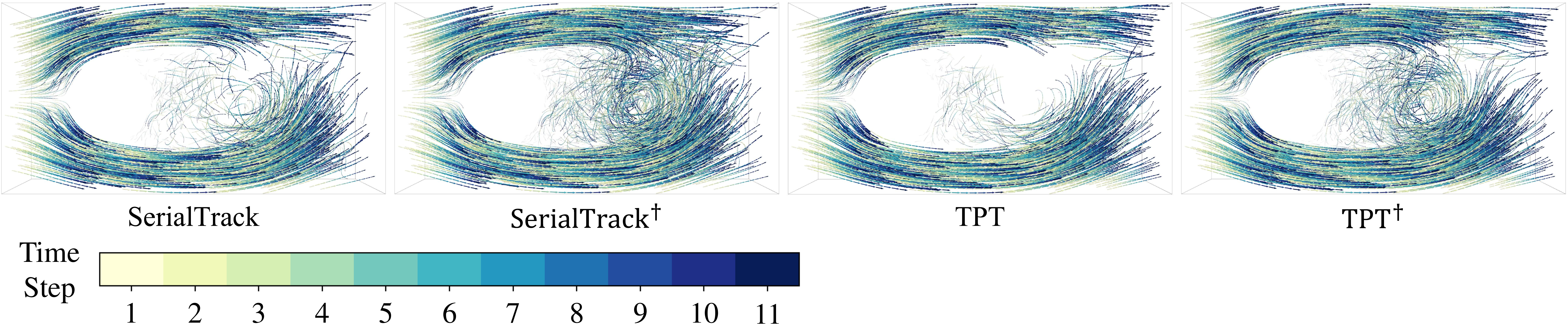}
 	\end{center}
	\vspace{4pt}
 	\caption{\textbf{Visualization of fully tracked trajectories in the CylinderFlow dataset.}}
 	\label{fig:cylinderflow_visual}
\end{figure}

As illustrated in Table \ref{tab:cylinderflow} and Fig. \ref{fig:cylinderflow}a, T-PT surpasses SerialTrack at a low particle density (\ie, $n_p=3,940$), with the performance disparity increasing by the final frame.
However, as the time interval increases, \ie, from $\Delta t=30$ to $\Delta t=90$, both T-PT and SerialTrack encounter a significant performance decline, with the disparity between them diminishing.
Conversely, SerialTrack exhibits superior performance at higher particle densities, \ie, $n_p=9,850/19,700$, and a larger time interval, \ie, $\Delta t=90$.
The integration of DiffTR-Flow3D into T-PT and SerialTrack results in a significant performance improvement across all settings.
At a low sampling rate ($\Delta t=90$), the integration of DiffTR-Flow3D increases the mean $E_y$ by 18.1-28.1 percentage points across all particle densities.
Under a high sampling rate ($\Delta t=30$), $\text{T-PT}^{\dag}$ sustains $E_y$ above 96\% and $\text{SerialTrack}^{\dag}$ above 85\% across varying particle densities after tracking over 34 consecutive frames.
This performance persistence highlights its superior capability in multi-frame particle tracking scenarios.
Qualitative examples of the resulting fully tracked trajectories are shown in Fig. \ref{fig:cylinderflow_visual}.

To demonstrate the benefits of incorporating trajectory information, we empirically show its performance improvement under large displacements (as shown in Fig. \ref{fig:cylinderflow}b).
Our experiments reveal that trajectory information improves tracking accuracy, with the gain growing as tracking proceeds over successive frames and with increasing particle density.
Notably, under $\Delta t=90$ and $n_p=19,700$, $\text{SerialTrack}^{\dag}$ improves the final-frame tracking accuracy by 23.83 percentage points over vanilla SerialTrack (73.96\% vs. 50.13\%).
These results underscore the superiority of our methodology that synergistically integrates temporal prediction with a learned diffusion framework.

\begin{figure}[!tbp]
 	\begin{center}
 	\includegraphics[width=1.0\columnwidth]{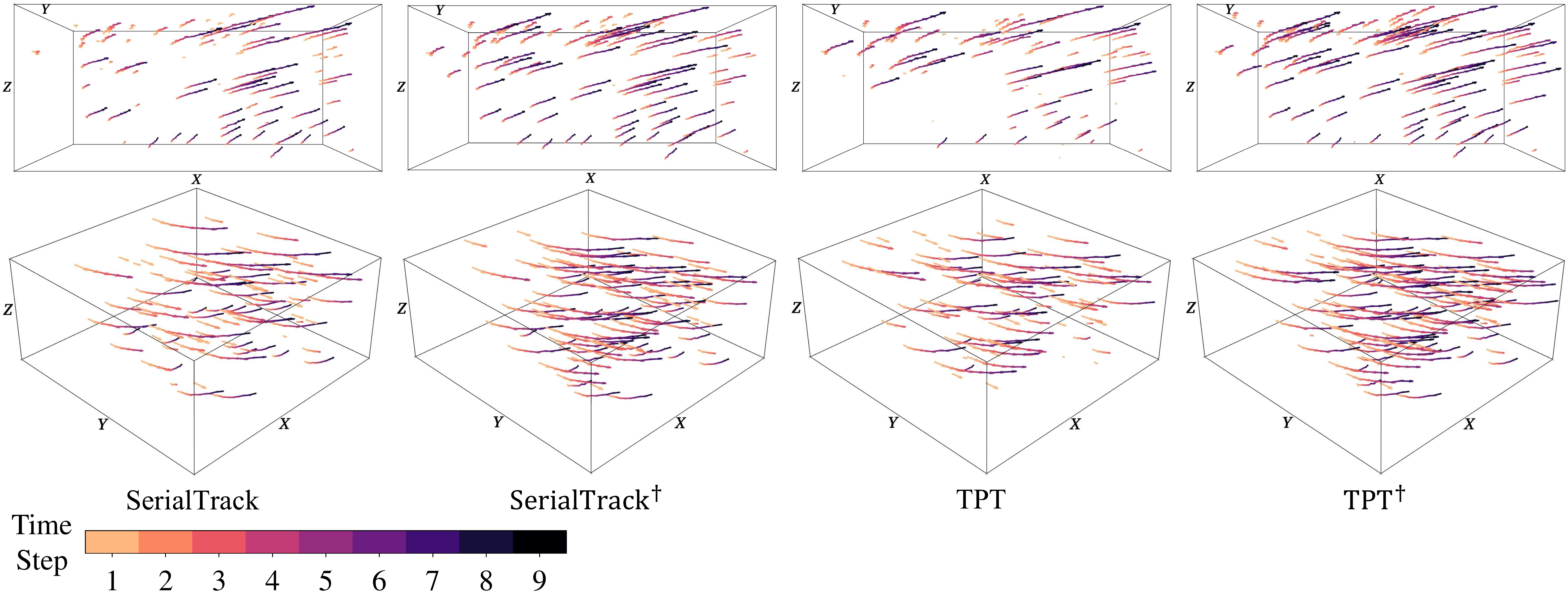}
 	\end{center}
	\vspace{4pt}
 	\caption{\textbf{DiffTR-Flow3D applied to the mechanical shear testing performed on hydrogels to characterize their viscoelastic behavior under shear deformation.}}
 	\label{fig:hydrogel_shear}
\end{figure}

To further investigate DiffTR-Flow3D's tracking capabilities specifically within relatively sparse scenarios, we apply the model to hydrogel shear testing for viscoelastic behavior characterization \cite{yang2022serialtrack}, which simulates biological applications such as traumatic brain injury \cite{bar2016strain}.
This experiment evaluates the complete processing pipeline from experimental confocal volumetric images to particle trajectories: particle positions are first extracted using SerialTrack3D's detection and localization routine and then linked using T-PT or SerialTrack, with or without DiffTR-Flow3D.
For each baseline and its enhanced counterpart, we use identical localization settings and extracted coordinates to assess the contribution of DiffTR-Flow3D downstream of experimental particle localization.
Fig. \ref{fig:hydrogel_shear} demonstrates that our method also improves trajectory retention in the 3D PTV pipelines.
For T-PT, the number of tracked particles sharply decreases from 88 to 21 during tracking, while integration with our method limits this decline to 106 to 64 particles across frames.
Similarly, SerialTrack shows enhanced performance when augmented with our method, maintaining 99 tracked particles in the first timestep (\vs 89 in vanilla) and 52 in the final (\vs 23 in vanilla).
The associations added by DiffTR-Flow3D show affine-motion residuals comparable to those shared with the original SerialTrack and T-PT pipelines (Appendix \ref{sec:supp-hydrogel-consistency}, Table \ref{tab:r1-hydrogel}), providing kinematic consistency evidence that complements the increased trajectory retention.

\par\vspace{12pt}
\section{Discussion and Conclusions}
In this work, we introduce DiffTR-Flow3D, a diffusion-based framework for multi-frame displacement estimation that supports particle linking under large inter-frame displacements.
DiffTR-Flow3D utilizes a powerful two-step flow motion estimator to handle diverse displacement distributions in particle dynamics:
it first coarsely places the source particles within the vicinity of the target position (\ie, flow motion estimation);
it then eliminates the residual position offset with soft alignment operations by capturing local and global correlations (\ie, flow motion refinement).
For multi-frame trajectory prediction,  DiffTR-Flow3D employs fast test-time fitting to extrapolate the next particle position in each trajectory from prior trajectory information.
These extrapolated positions serve as the initial conditions for the diffusion-based estimator.
A multi-frame dataset named TR-Flow3D is synthesized to mimic real-world conditions as closely as possible, including various fluid dynamics scenarios across broad observation volumes and time periods, which can serve as a valuable resource to foster more extensive exploration in this field.
DiffTR-Flow3D yields the best performance on the FluidFlow3D dataset and the newly created TR-Flow3D dataset, outperforming competing methods by substantial margins under both double-frame and multi-frame settings.
The robustness of DiffTR-Flow3D is also verified through experiments under varying conditions.
On CylinderFlow, integrating DiffTR-Flow3D with T-PT and SerialTrack improves long-term association across different particle densities and sampling intervals; on the experimental hydrogel case, it substantially increases the number of trajectories retained through the final frame.
These results demonstrate the out-of-distribution generalization capabilities of DiffTR-Flow3D.

A primary limitation of this framework is that its scope is restricted to large-displacement, temporally under-sampled regimes.
These conditions typically arise when hardware constraints or high-speed flow dynamics result in scenarios where particle displacements between consecutive frames are excessively large for standard matching methods.
For instance, this challenge is particularly acute in sparse and under-sampled observations (\eg, traumatic brain injury \cite{bar2016strain}).
However, applications involving high temporal resolution and small particle displacements (\eg, microfluidic dynamics \cite{cierpka2012particle} or vortex evolution studies \cite{wolf2011time}) prioritize particle reconstruction, because particle matching is generally more straightforward under such conditions.
For such scenarios, STB \cite{schanz2016shake,Jahn2021} remains the preferred choice due to its superior ability to reconstruct particles, even when the particle density is high.
Thus, developing noise-robust particle reconstruction methodologies capable of resolving long-range temporal dependencies remains a promising future direction.
Another limitation is that DiffTR-Flow3D focuses on estimating inter-frame displacements for particle association, rather than resolving the continuous, sub-frame flow evolution between observed frames.
Consequently, accurately extracting trajectory statistics and Lagrangian quantities that rely on higher-order temporal derivatives, such as energy spectra and turbulence statistics, remains challenging.
Finally, the current implementation is also limited by its memory requirements for large particle sets, as the storage required by the brute-force neighbor search and dense global attention grows quadratically with particle number (Appendix~A.3).
Memory-efficient implementations and overlapping sub-volume processing could be explored to improve scalability.

Our study, which focuses on displacement estimation for multi-frame particle tracking, takes a step toward maintaining trajectories in temporally under-sampled sequences.
Building upon this foundation, several promising avenues emerge for future exploration.
First, explicitly integrating the physical laws of fluid dynamics into neural networks remains a compelling direction.
For instance, \cite{jin2021nsfnets} employs physics-informed neural networks (PINNs) to simulate incompressible flows by solving Navier-Stokes equations, while \cite{clark2023reconstructing} leverages PINNs to reconstruct Eulerian grids from sparse particle trajectories.
Once trained, PINNs enable low-cost inference for applications with recurring flow patterns.
However, their reliance on substantial ground-truth data limits applicability in data-scarce scenarios, which is still an open challenge in multi-frame particle tracking.
A viable remedy lies in leveraging semi-supervised or unsupervised approaches to utilize abundant available unlabeled data in data-scarce scenarios.
Second, a promising direction is to formulate 3D PTV as a point tracking task \cite{doersch2022tap}, which extends optical flow estimation to longer timeframes with occlusion estimation. However, adapting algorithms originally designed for dense and rich pixel contexts to Eulerian grids or even Lagrangian particles remains a significant challenge.

\section{Methods}\label{sec4}
\subsection{Preliminary}
The key idea of the diffusion model is formulated as a parameterized Markov chain, which iteratively destroys structure in a data distribution $d_0 \sim q(d)$ over $N$ steps, ultimately to a Gaussian distribution, \ie, the diffusion process $q(d_{1:N} | d_{0})$. Then, a reverse process $p_{\theta}(d_{n-1} | d_{n})$ is learned to restore structure in the data by eliminating the noise.

The diffusion process is formulated as a fixed Markov chain that systematically introduces Gaussian noise to the data, guided by a predefined variance schedule $\beta_{1:N}$:
\begin{equation}
q(d_{1:N}|d_{0}) = \prod\nolimits_{n=1}^{N} q(d_{n}|d_{n-1}),
\end{equation}
\begin{equation}
q(d_{n}|d_{n-1}) = \mathcal{N}(d_{n};\sqrt{1-\beta_{n}}d_{n-1},\beta_{n}\bm{I}),
\end{equation}
where $\bm{I}$ is the identity matrix. By defining $\alpha := 1 - \beta$ and $\bar{\alpha}_{n} := \prod_{s=1}^{n} \alpha_{s}$, the forward process allows for closed-form sampling of $d_n$ at an arbitrary timestep $n$ with a reparameterization trick \cite{kingma2013auto}:
\begin{equation}
\label{eq:diff}
q(d_{n}|d_{0}) = \mathcal{N}(d_{n};\sqrt{\bar{\alpha}_{n}}d_{0},(1-\bar{\alpha}_{n})\bm{I}).
\end{equation}
The reverse process is also a Markov chain with learnable parameters $\theta$ to recover the original data $d_0$ from the noise data $d_n$: \par\vspace{-\baselineskip}
\begin{align}
    p_{\theta}(d_{0:N}) &= p_{\theta}(d_{N}) \prod\nolimits_{n=1}^{N} p_{\theta}(d_{n-1}|d_{n}), \\
    p_{\theta}(d_{n-1}|d_{n}) &= \mathcal{N}(d_{n-1}; \mu_{\theta}(d_{n}, d_{0}), \sigma_n^2\mathbf{I}), \label{eq:5}
\end{align}
where $\mu_{\theta}$ denotes the predicted mean of the parameterized neural network and $\sigma_n^2$ denotes the pre-defined variance.
Following \cite{ho2020denoising}, the mean $\mu_{\theta}$ is reparameterized via a noise prediction network $\epsilon_{\theta}(d_n, n, c_f)$.
Thus, the training objective can be simplified as predicting the noise $\epsilon$ added to the data:
\begin{equation}
     \mathcal{L}_{\theta} = \mathbb{E}_{n, d_0, c_f, \epsilon \sim \mathcal{N}(0,\bm{I})} [\vert\vert \epsilon - \epsilon_{\theta}(d_n, n, c_f) \vert\vert^{2}_{2}],
\end{equation}
where $c_f$ is the task-specific condition.
In our problem setting, $c_f$ represents the source point set $P_s$ and the target point set $P_t$ for flow estimation.
For computational efficiency, we adopt DDIM \cite{song2020denoising} as our sampling strategy to estimate the flow motion.

\subsection{DiffTR-Flow3D}
Particles in multi-frame particle tracking commonly exhibit significant velocity fluctuations, resulting in displacements across various scales.
This complexity poses a substantial challenge to achieving precise flow estimation in a single step.
Following previous recurrent methods \cite{wei2021pv,liang2023recurrent}, our approach also employs a coarse-to-fine structure, but it is built upon a GMSF-based paradigm \cite{zhang2024gmsf} without iterative updates for flow estimation.
Notably, apart from the use of historical trajectory priors, DiffSF \cite{zhang2024diffsf} is the method most closely related to ours, as both build on GMSF \cite{zhang2024gmsf} and incorporate a conditional diffusion process.
Unlike DiffSF\cite{zhang2024diffsf}, which estimates the full displacement in both stages, our method performs explicit coarse-to-fine refinement by predicting a residual correction and adding it to the coarse displacement, a design intended for large inter-frame displacements (Tables 1-2).

\noindent\textbf{Step 1: Flow motion estimation.} First, with the input particle sets $\{P_s\!\in\!\mathbb{R}^{N_s \times 3}, P_t\!\in\!\mathbb{R}^{N_t \times 3}\}$, an initial estimation step is introduced to roughly place the source particles within the vicinity of the target positions.

\textit{Feature extraction.}
As stated in \cite{schroder20233d}, flow dynamics and coherent structures can be fundamentally characterized as an ensemble of discrete fluid elements traversing within the local flow field.
Inspired by this, we construct a comprehensive descriptor characterizing detailed topological information for each discrete particle $x_i$ by selecting the $k$ nearest neighbor particles $x_j \in \mathcal{N}(i)$ within the same timestep.
Specifically, we project a three-dimensional coordinate $x_i$ into a higher-dimensional feature space $\mathbb{R}^{D}$ conditioned on its neighbors $x_j \in \mathcal{N}(i)$:
\begin{equation}
x_i^e = \max_{x_j \in \mathcal{N}(i)} \phi_e(x_i, x_j-x_i),
\label{eq:7}
\end{equation}
where $\phi_e\!:\!\mathbb{R}^{6} \to \mathbb{R}^{C}$ denotes a lifting operation implemented as a sequence of linear layers, batch normalization, and LeakyReLU activations, following DGCNN \cite{wang2019dynamic}.
The global max pooling operator $\max$ is utilized to aggregate point features while maintaining permutation invariance.
To build a feature pyramid with semantics from low to high levels, multiple layers are stacked together to get the final feature descriptors.
Finally, for the source point set $P_s$ and the target point set $P_t$, we can respectively get context feature descriptors denoted as $F_s^e=\{x_i^e\}^{N_s}_{i=1}\!\in\!\mathbb{R}^{N_s \times C}$ and $F_t^e=\{x_j^e\}^{N_t}_{j=1}\!\in\!\mathbb{R}^{N_t \times C}$.

\textit{Motion estimation.} The intuitive formulation of motion estimation is as follows: first, the target points are aligned with the source points by taking the location that gives the highest correlation; then, the flow motion can be obtained by computing the difference between these two point sets.
The core issue is how to achieve point alignment in a differentiable manner.
Previous methods like FLOT \cite{puy2020flot} and GotFlow3D \cite{liang2023recurrent} perform point matching by solving the optimal transport problem.
In contrast, we directly adopt global feature similarity matrices between the source and the target descriptors as correspondences for flow motion estimation. This process can be implemented using a simple attention operator:
\begin{equation}
D_c^e = \text{softmax}(\frac{F_s^e {F_t^e}^\top}{\sqrt{d}})P_t - P_s,
\label{eq:8}
\end{equation}
where $D_c^e\!\in\!\mathbb{R}^{N_s \times 3}$ denotes the flow from source to target points and $d$ is the dimension of $F_s^e$.

In realistic measurements, the particle mismatches can be amplified by optical distortion, overlapping particle images, or out-of-plane particle motion, which are ubiquitous and inevitable.
A prevalent approach is to directly filter these out by employing a pre-defined threshold \cite{schanz2016shake,yang2022serialtrack}, which heavily relies on hand-crafted rules and exhibits poor generalization.
Moreover, during multi-frame tracking, these particles may only be temporarily obscured in just one or two frames.
The direct removal of them would result in a continuous decline in the quantity of observed measurements.

In light of this observation, we endeavor to generate a flow approximation for missing particles, leveraging the accurately computed scene flow estimations from their adjacent non-occluded areas, as:
\begin{equation}
D_{coarse} = \text{softmax}(\frac{(W_q F_s^e)(W_k F_s^e)^\top}{\sqrt{d}})D_c^e,
\label{eq:9}
\end{equation}
where $D_{coarse}\!\in\!\mathbb{R}^{N_s \times 3}$ denotes the flow approximation, and $W_q\!\in\!\mathbb{R}^{C \times C}$, $W_k\!\in\!\mathbb{R}^{C \times C}$ are learnable embedding matrices.

\noindent\textbf{Step 2: Flow motion refinement.}
Next, given the input particle sets $\{P_s\!\in\!\mathbb{R}^{N_s \times 3}, P_t\!\in\!\mathbb{R}^{N_t \times 3}\}$ and the coarse flow estimation $D_{coarse}\!\in\!\mathbb{R}^{N_s \times 3}$, we update the input particle sets to $\{\hat{P}_s\!\in\!\mathbb{R}^{N_s \times 3}, P_t\!\in\!\mathbb{R}^{N_t \times 3}\}$, where $\hat{P}_s=P_s+D_{coarse}$.
Following this, we extract the context features of these new input particle sets in a manner analogous to \textbf{Step 1}, obtaining context feature descriptors $F_s^r=\{x_i^r\}^{N_s}_{i=1}\!\in\!\mathbb{R}^{N_s \times C}$ and $F_t^r=\{x_j^r\}^{N_t}_{j=1}\!\in\!\mathbb{R}^{N_t \times C}$ via $\phi_r$, which also represents a sequence of linear layers.
Building upon these descriptors, we further enhance them by employing soft alignment operations. This is achieved through dot-product attention mechanisms designed to capture both local and global correlations. Specifically, the enhanced features are computed as follows:

\textit{Local correlation.}
The self-attention mechanism, essential to the Transformer architecture, is fundamentally a set operator, making it an ideal approach for constructing correspondences for unordered point clouds.
Therefore, we adopt a Point Transformer \cite{zhao2021point} to enhance the context descriptors $x_i^r$ by confining the computation of soft attention to the local neighborhood of each particle, \ie, $x_l^r \in \mathcal{N}_r(i)$:
\begin{equation}
\hat{x}_i^r = \sum\nolimits_{x_l^r \in \mathcal{N}_r(i)}\!\!\!\text{softmax}(\gamma(\phi_q(x_i^r)-\phi_k(x_l^r)+\delta)) \odot (\phi_v(x_l^r)+\delta),
\label{eq:10}
\end{equation}
where $\mathcal{N}_r(i)$ denotes the $k$ nearest-neighbor set of $x_i^r$ in the refined point set, $\phi_q, \phi_k, \phi_v: \mathbb{R}^{C}\!\to\!\mathbb{R}^{C}$ are linear projections to perform pointwise feature transformations, $\gamma: \mathbb{R}^{C}\!\to\!\mathbb{R}^{C}$ is a mapping function to produce attention vectors, and $\delta\!\in\!\mathbb{R}^{C}$ is the relative position embedding derived from the coordinates of $x_i^r$ via a linear layer; all of these are trainable parameters.
This process is concurrently applied to $F_s^r$ and $F_t^r$, ultimately yielding the refined feature descriptors, denoted as $\hat{F}_s^r=\{\hat{x}_i^r\}^{N_s}_{i=1}\!\in\!\mathbb{R}^{N_s \times C}$ and $\hat{F}_t^r=\{\hat{x}_j^r\}^{N_t}_{j=1}\!\in\!\mathbb{R}^{N_t \times C}$.

\textit{Global correlation.}
To estimate particle motion using global correlations, the correlated dependencies between feature descriptors $\hat{F}_s^r\!\in\!\mathbb{R}^{N_s \times C}$ and $\hat{F}_t^r\!\in\!\mathbb{R}^{N_t \times C}$, which are extracted independently, are taken into account.
In detail, several Transformer-like blocks are built by alternately stacking self-attention, cross-attentions and linear layers to calculate correlation features:
\begin{equation}
\tilde{F}_s^p = \phi_\mathcal{T}(\hat{F}_s^r,\hat{F}_t^r), ~~~\tilde{F}_t^p = \phi_\mathcal{T}(\hat{F}_t^r,\hat{F}_s^r),
\label{eq:11}
\end{equation}
where $\phi_\mathcal{T}$ denotes a standard Transformer block \cite{vaswani2017attention} (detailed in Table \ref{tab:para_config}, Appendix \ref{secB}) with the first input as the query and the second as the key and value.
These descriptors are then fed into the aforementioned motion estimation and approximation processes, yielding the residual correction relative to $\hat{P}_s$, denoted as $D_{pred}\in\mathbb{R}^{N_s\times 3}$.

\noindent\textbf{Diffusion-based training and inference.}
DiffTR-Flow3D is trained using a diffusion paradigm. We simulate the process by defining a trajectory from the ground-truth flow to Gaussian noise, controlled by $\bar{\alpha}_{n}$ (Eq. \ref{eq:diff}).
Given the source $P_s \in \mathbb{R}^{N_s \times 3}$ and the target $P_t \in \mathbb{R}^{N_t \times 3}$ as conditions, the model takes the noisy flow $D_n \in \mathbb{R}^{N_s \times 3}$ as input to predict a residual flow component $D_{pred} \in \mathbb{R}^{N_s \times 3}$, which is then added to $D_n + D_{coarse}$ to obtain the clean flow $\hat{D}_{s \rightarrow t} \in \mathbb{R}^{N_s \times 3}$.
Notably, this noisy flow is added to $P_s$ during \textbf{Step 1} to simulate particle displacement under advection (Fig. \ref{fig:method}a), thereby enabling DiffTR-Flow3D to learn noise-robust displacement correction patterns.
The detailed procedure is outlined in Algorithm \ref{alg:training} (Appendix \ref{secB}).

DiffTR-Flow3D is trained end-to-end on two NVIDIA RTX 4090 GPUs for 500 epochs using a OneCycle learning rate schedule \cite{smith2018disciplined} with a batch size of 8, taking approximately 100 hours.
We utilize the AdamW optimizer with a learning rate of $4 \times 10^{-5}$ and a weight decay of $1 \times 10^{-4}$. The number of diffusion timesteps is set to 20.
We apply a robust regression loss to supervise this process, as follows:
\begin{equation}
\mathcal{L} = \sum\nolimits_{i}(\lVert\hat{D}_{s \rightarrow t}(i)-D_{gt}(i)\rVert_{1} + \delta)^q,
\label{eq:12}
\end{equation}
where $D_{gt}$ represents the ground-truth flow and  $\lVert \cdot \rVert_{1}$ computes the $L_1$ norm. Following \cite{liu2022camliflow,zhang2024gmsf}, $\delta$ is chosen as $0.01$ and $q$ is set to $0.4$.

For inference, DiffTR-Flow3D performs the reverse process to estimate the displacement field between observed frames.
This procedure does not explicitly solve the transport equation over time; rather, it iteratively denoises an initial noisy displacement toward an estimate that aligns the source points with their observed targets.
To facilitate efficient denoising sampling, we employ a cosine noise schedule and leverage DDIM acceleration to condense this process into two steps.

\noindent\textbf{Multi-frame trajectory prediction.}
Beyond double-frame flow estimation, our goal is to track the motion of particles over an extended period of time.
For efficiency, we first build a powerful double-frame flow motion estimator based on the aforementioned method, and then perform a fast test-time fit of a ridge model to capture the consistency of spatiotemporal particle trajectories, as shown in Fig. \ref{fig:method}e.
The trajectories of particles are a fundamental characteristic of flow dynamics, exhibiting pronounced temporal correlations.
For instance, \cite{schanz2013shake} uses a Savitzky-Golay filter to fit the data points and extrapolate their positions, whereas  \cite{schanz2016shake} opts for a more flexible and reliable Wiener filter.
However, these algorithms necessitate careful calibration and parameter selection to strike an optimal balance between noise suppression and signal preservation.
In contrast, we prioritize robust and generalizable, albeit coarse-grained, predictions to provide reliable initial conditions for the reverse diffusion process.
In light of this, we simply learn a polynomial function to fit particle trajectories by optimizing a ridge regression model:
\begin{equation}
\min_{W} \{ \lVert XW - Y \rVert^2_2 + \alpha \lVert W \rVert^2_2 \},
\label{eq:13}
\end{equation}
where $\lVert \cdot \rVert$ denotes the Euclidean norm and $\alpha$ is the regularization parameter.
$X \in \mathbb{R}^{N_k \times M}$ is the polynomial features derived from $N_k$ timesteps, $Y \in \mathbb{R}^{N_k \times N_p \times 3}$ is the particle positions over these timesteps, and $W \in \mathbb{R}^{M \times N_p \times 3}$ is the learnable weight matrix.
Here, $N_k$ is the number of particle frames, $N_p$ is the number of particles being tracked, and $M$ is the degree of the polynomial features.

\subsection{Evaluation metrics} \label{sec:eval}
To quantitatively evaluate and compare the performance of all methods, we employ the evaluation metrics used in \cite{puy2020flot,wei2021pv}, including EPE and scalable metrics in non-dimensional forms, \ie, ACC Strict, ACC Relax, and Outliers.
Let $\hat{f}_{e}$ denote the estimated flow, and $f_{gt}$ be the ground-truth flow. The evaluation metrics are defined as follows:
\begin{itemize}
\item \textbf{EPE}: the average end-point error in meters, \ie, $\frac{1}{N}\sum^{N}_{i=1} \vert\vert \hat{f}_{e}^i - f_{gt}^i \vert\vert_{2}$.
\item \textbf{ACC Strict}: the percentage of points with a relative error less than 5\%.
\item \textbf{ACC Relax}: the percentage of points with a relative error less than 10\%.
\item \textbf{Outliers}: the percentage of points with a relative error larger than 10\%.
\end{itemize}

\noindent To facilitate comparison with particle tracking algorithms, we define tracking accuracy as the proportion of correctly tracked trajectories during the multi-frame tracking process:
\begin{equation}
E_y = \frac{N_p}{N_g},
\label{eq:tracking_acc}
\end{equation}
where $N_p$ denotes the number of correctly tracked trajectories and $N_g$ indicates the total number of ground-truth trajectories.

\subsection{Synthetic datasets}
To satisfy the supervised learning strategy for our method, we create a synthetic multi-frame 3D fluid flow dataset, named TR-Flow3D, comprising 19,500 training samples and 1,950 test samples.
In this work, we adhere to the following procedures:\\
\noindent\textbf{Particle set initialization.}
First, we directly sample a set of $2,048$ particles that are randomly seeded within the predefined three-dimensional observation volume (as shown in Table \ref{tab:dataset}).
Following the configurations in \cite{liang2023recurrent}, we set the observation volume and the inter-frame sampling interval $\Delta t$ accordingly.
To showcase the heterogeneity of our data, we utilize the particle displacement ratio $\rho$, a standard metric in the field of Particle Tracking Velocimetry (PTV) \cite{malik1993particle}.
The metric is formulated as:
\begin{equation}
\rho = \frac{d_a}{u_{\max}\Delta t},
\label{particle_displacement_ratio}
\end{equation}
where $d_a$ represents the mean distance between particles (reflecting particle density), and $u_{\max}\Delta t$ corresponds to the maximum particle displacement between consecutive frames. As evidenced in Table \ref{tab:dataset}, our dataset covers a wide spectrum of scenarios, exhibiting particle displacement ratios ranging from 0.23 to 8.75.
Notably, 66.84\% of samples exhibit $\rho<1$, meaning that the maximum displacement exceeds the mean inter-particle spacing and therefore creates substantial matching ambiguity.

\begin{table}[t]
\setlength{\tabcolsep}{3pt}
\caption{\textbf{Generation condition for diverse flows in the TR-Flow3D dataset.}}
\begin{tabular}{@{}llllcc@{}}
\toprule
Flow field pattern& Observation Volume& $\Delta t$& $\rho$& $\text{Train Num.}$& $\text{Test Num.}$\\
\midrule
\makecell[l]{Channel\\flow}& $\{(0.25\pi, 0.5, 0.25\pi)\}$& \makecell[l]{$\{0.03, 0.04, 0.05,$ \\~~\!$0.06, 0.07, 0.08\}$}& $[0.41, 4.30]$& $3000$& $300$\\
\midrule
\makecell[l]{Transitional\\boundary\\flow}& \makecell[l]{$\{(0.5\pi, 1.0, 0.5\pi),$ \\~~\!$(\pi, 2.0, \pi)\}$}& $\{0.12, 0.16, 0.2\}$& $[0.28, 3.25]$& $3000$& $300$\\
\midrule
\makecell[l]{Forced\\MHD\\turbulence}& \makecell[l]{$\{(0.25\pi, 0.25\pi, 0.25\pi),$ \\~~\!$(0.5\pi, 0.5\pi, 0.5\pi)\}$}& $\{0.03, 0.04, 0.05\}$&  $[0.46, 3.68]$& $3000$& $300$\\
\midrule
\makecell[l]{Forced\\isotropic\\turbulence}& \makecell[l]{$\{(0.25\pi, 0.25\pi, 0.25\pi),$ \\~~\!$(0.5\pi, 0.5\pi, 0.5\pi)\}$}& $\{0.06, 0.07, 0.08\}$& $[0.28, 3.25]$& $3000$& $300$\\
\midrule
\makecell[l]{Uniform\\flow}& $\{(0.25\pi, 0.25\pi, 0.25\pi)\}$& $\{0.03, 0.04, 0.05\}$& $[0.52, 8.75]$& $3000$& $300$\\
\midrule
\makecell[l]{Beltrami\\flow}& $\{(2.0, 2.0, 2.0)\}$& \makecell[l]{$\{[0.9, 1.9],$ \\~~\!$[2.1, 3.1],$ \\~~\!$[3.3, 4.3]\}$}& $[0.23, 0.88]$& $4500$& $450$\\
\botrule
\end{tabular}
We follow the generation conditions outlined in \cite{liang2023recurrent}, generating Uniform and Beltrami flows using computational fluid dynamics (CFD), and extracting other flow patterns from the Johns Hopkins Turbulence Database.
We further extend this setup to a multi-frame setting, where each sample consists of ten particle frames and nine corresponding ground-truth flow fields.
TR-Flow3D is generated through an exhaustive combination of parameters listed in this table, including flow motion patterns, the range of observation volume, and inter-frame sampling interval $\Delta t$, ultimately yielding 19,500 training samples and 1,950 test samples. $\rho$ denotes the particle displacement ratio (see Eq. \ref{particle_displacement_ratio} for the definition).
All particle sets are normalized to the cubic domain $[0, 2\pi]^{3}$ before neural network processing.
\label{tab:dataset}
\end{table}

\noindent\textbf{Flow fields simulation.}
Based on the initial particle set and given time intervals, we continuously obtain new coordinate positions of the particles through CFD simulations by solving the Navier-Stokes equations. Using the updated particle set, we repeat this process until we acquire a sufficient number of particle frames.
We implement particle tracking using the built-in functions from the Johns Hopkins Turbulent Database (JHTDB), \ie, \textit{GetPosition}, which tracks a list of particles simultaneously and returns their final locations at the end of the user-specified trajectory integration time. Particle tracking is performed using a second-order Runge-Kutta integration scheme for time and fourth-order Lagrange interpolation for space to request CFD data across various flow regimes, including Forced isotropic turbulence, Forced MHD turbulence, Turbulent channel flow, and Transitional boundary layer.
In addition, we simulate unsteady three-dimensional Beltrami flow based on the analytical solutions proposed by Ethier and Steinman \cite{ethier1994exact}, and include uniform flow to enhance the diversity of the data. Ultimately, we generate a synthetic dataset comprising 19,500 training samples and 1,950 test samples, with each sample containing 10 consecutive frames and 9 corresponding flows between adjacent frames. This results in a total of 214,500 particle frames and 193,050 generated flows.

\noindent\textbf{Particle loss simulation.}
The practical utility of 3D PTV has historically been constrained by particle reconstruction errors.
To enhance the realism of the simulated training data, we replace 1\% of the particle points with points drawn from a Gaussian distribution within the predefined 3D observation volume.
For robustness evaluation, we vary this replacement ratio at test time as described in Section \ref{sec3}.

\noindent\textbf{Data preprocessing.}
To facilitate stable training, all particle sets are normalized to a unified cubic domain $[0, 2 \pi]^3$.
These coordinates are subsequently denormalized to their original physical scales for evaluation.
To improve generalization, snapshots from all flow topologies are aggregated into a single pool and stochastically shuffled, ensuring that each training batch contains a diverse mixture of flow types.

\subsection{Public datasets}
\noindent\textbf{FluidFlow3D} \cite{liang2023recurrent}. It is a synthetic collection of 3D particle sets and corresponding flow fields, sourced from the JHTDB or generated by computational fluid dynamics (CFD). It contains 16,000 training samples across six flow scenarios and four test datasets that assess the influence of particle density, out-of-view particles, and coordinate offsets.

\noindent\textbf{CylinderFlow} \cite{KHOJASTEH2022107725}. It is constructed by capturing the wake flow downstream of a smooth cylinder at a Reynolds number of 3900, including Eulerian velocities and Lagrangian particle trajectories obtained via Direct Numerical Simulation (DNS) with the Incompact3d solver and a fourth-order Runge-Kutta scheme.

\section{Data availability}
The synthetic TR-Flow3D dataset can be accessed at \url{https://doi.org/10.6084/m9.figshare.27617541}.
The open-source FluidFlow3D dataset can be downloaded at \url{https://github.com/JiamingSkGrey/FluidFlow3D-family}.
The open-source CylinderFlow dataset can be downloaded at \url{https://doi.org/10.15454/GLNRHK}.
The open-source experimental test case of hydrogel shear can be downloaded at \url{https://minds.wisconsin.edu/handle/1793/82901}.

\section{Code availability}
The source code is available at \url{https://github.com/weijianan1/DiffTR-Flow3D}.
The implementation of T-PT algorithm adopted in this paper is available at \url{https://github.com/FranckLab/T-PT}.
The implementation of SerialTrack algorithm adopted in this paper is available at \url{https://github.com/FranckLab/SerialTrack}.
The implementation of STB algorithm adopted in this paper is available at \url{https://github.com/JHU-NI-LAB/OpenLPT_Shake-The-Box}.

\backmatter

\bibliography{main}%

\clearpage
\begin{appendices}

\setcounter{figure}{0}
\setcounter{table}{0}
\renewcommand{\theHfigure}{\thesection.\arabic{figure}}
\renewcommand{\theHtable}{\thesection.\arabic{table}}
\renewcommand{\theHalgorithm}{\thesection.\arabic{algorithm}}
\renewcommand{\thefigure}{\thesection.\arabic{figure}}
\renewcommand{\thetable}{\thesection.\arabic{table}}
\renewcommand{\thealgorithm}{\thesection.\arabic{algorithm}}

\renewcommand{\tablename}{Table}
\renewcommand{\figurename}{Fig.}
\floatname{algorithm}{Algorithm}

\section{More experimental results}\label{secA}

\subsection{Ablation study of hyperparameters in DiffTR-Flow3D}
DiffTR-Flow3D incorporates a variety of hyperparameters inherited from the field of optical flow estimation.
The majority of these parameters strictly follow the configurations established by GMFlow \cite{xu2022gmflow} and GMSF \cite{zhang2024gmsf} for the backbone and matching modules.
Here, we conduct ablation experiments to evaluate the impact of the hidden dimension size, \ie, the dimensionality to which particle coordinates are projected, as well as the number of training and DDIM sampling steps utilized during the diffusion process.
As illustrated in Table \ref{tab:ablation_hidden}, variations in the hidden dimension size lead to only marginal performance changes. The hidden dimension of 128 provides the best overall trade-off, achieving the highest ACC Strict and ACC Relax and the lowest Outliers, despite a slightly higher EPE than the 64-dimensional setting. Therefore, we adopt 128 as the default hidden dimension.
Furthermore, as demonstrated in Table \ref{tab:ablation_diffusion_strategy}, the model's performance drops significantly without the diffusion strategy. We attribute this to the difficulty of mapping the input directly to the complex target distribution without iterative refinement. Equipping the base model with a diffusion process yields substantial performance gains, with the optimal configuration achieved at 20 training steps and 2 sampling steps.

\begin{table*}[h]
\centering
\caption{\textbf{Ablation on different hidden dimension sizes on the TR-Flow3D dataset under the multi-frame setting.} The best metric values are shown in bold.}\label{tab:ablation_hidden}%
\begin{tabular}{@{}llllll@{}}
\toprule
Hidden Dimension & EPE$\downarrow$ & ACC Strict$\uparrow$ & ACC Relax$\uparrow$ & Outliers$\downarrow$ \\
\midrule
64               & \textbf{0.00313} & 98.15\% & 98.45\% & 1.55\%  \\
128     			& 0.00316& \textbf{98.48\%}& \textbf{98.72\%}& \textbf{1.28\%} \\
256     			& 0.00405 & 98.24\% & 98.61\% & 1.39\%  \\
512              & 0.00570 & 97.41\% & 97.93\% & 2.07\%  \\
\botrule
\end{tabular}
\end{table*}

\begin{table*}[h]
\centering
\caption{\textbf{Ablation study of the diffusion strategy and diffusion steps on the TR-Flow3D dataset under the multi-frame setting.} The default setting is shown in bold.}
\label{tab:ablation_diffusion_strategy}
\small
\setlength{\tabcolsep}{4pt}
\begin{tabular}{@{}lllllll@{}}
\toprule
\shortstack[l]{Training\\Strategy} & \shortstack[l]{Training\\Steps} & \shortstack[l]{Sampling\\Steps} & EPE$\downarrow$ & ACC Strict$\uparrow$ & ACC Relax$\uparrow$ & Outliers$\downarrow$ \\
\midrule
w/o diffusion & -- & -- & 0.00978 & 94.47\% & 95.24\% & 4.76\% \\
\midrule
Diffusion & 3 & 1	& 0.00553 & 96.74\% & 97.21\% & 2.79\%  \\
Diffusion & 5 & 1	& 0.00407 & 97.40\% & 98.04\% & 1.96\%  \\
Diffusion & 5 & 2	& 0.00353 & 98.31\% & 98.66\% & 1.34\%  \\
Diffusion & 20 & 1	& 0.00324 & 98.44\%& 98.71\%& 1.29\% \\
\textbf{Diffusion} & \textbf{20} & \textbf{2}	& \textbf{0.00316} & \textbf{98.48\%}& \textbf{98.72\%}& \textbf{1.28\%} \\
Diffusion & 20 & 5	& 0.00316 & 98.47\% & 98.72\% & 1.28\%  \\
\botrule
\end{tabular}
\end{table*}

\subsection{Module-level ablation study}\label{secA_module}
To further clarify the role of each module in the two-stage flow motion estimator (Section 5.2), we conduct a module-level ablation on the TR-Flow3D dataset under the multi-frame setting. All variants are trained with the same protocol as the full model.
(i) \textit{w/o refinement stage} outputs the coarse estimate $D_{coarse}$ of Step 1 directly, examining whether a single global matching step is sufficient for precise displacement estimation;
(ii) \textit{w/o coarse stage} removes Step 1 and applies the refinement stage to the original source points ($\hat{P}_s = P_s$), examining the role of pre-aligning source particles to the vicinity of their targets before refinement;
(iii) \textit{w/o local correlation} removes the Point Transformer layer (Eq. 10), which aggregates intra-frame neighbourhood information so that each descriptor encodes the local particle arrangement;
(iv) \textit{w/o global correlation} removes the Transformer blocks (Eq. 11), which exchange inter-frame information between all source and target particles before matching; and
(v) \textit{w/o local \& global} removes both modules.
As shown in Table~\ref{tab:ablation_module}, both stages are necessary.
Without the refinement stage, the EPE rises from 0.00316 to 0.10815, indicating that a single global matching step cannot resolve precise particle positions.
Removing the coarse stage also degrades performance (EPE 0.02529), showing that pre-aligning source particles near their targets is required for accurate refinement under large displacements.
Within the refinement stage, global correlation is indispensable (EPE 0.08352 without it), whereas removing local correlation causes a smaller but consistent degradation (EPE 0.00385).
Removing both modules performs on par with removing global correlation alone (EPE 0.07840), suggesting that local correlation improves accuracy only on top of inter-frame information exchange.

\begin{table}[h]
\centering
\caption{\textbf{Module-level ablation of DiffTR-Flow3D on the TR-Flow3D dataset under the multi-frame setting.} The full model is shown in bold.}
\label{tab:ablation_module}
\small
\begin{tabular}{@{}lcccc@{}}
\toprule
Variant & EPE$\downarrow$ & ACC Strict$\uparrow$ & ACC Relax$\uparrow$ & Outliers$\downarrow$ \\
\midrule
w/o refinement stage  & 0.10815 & 21.34\% & 37.78\% & 62.22\% \\
w/o coarse stage & 0.02529 & 84.59\% & 90.47\% & 9.53\% \\
\midrule
w/o local \& global & 0.07840 & 23.87\% & 42.85\% & 57.15\% \\
w/o global correlation & 0.08352 & 26.42\% & 41.81\% & 58.19\% \\
w/o local correlation  & 0.00385 & 97.74\% & 98.09\% & 1.91\% \\
\midrule
\textbf{Full model}    & \textbf{0.00316} & \textbf{98.48\%} & \textbf{98.72\%} & \textbf{1.28\%} \\
\bottomrule
\end{tabular}
\end{table}

\subsection{Efficiency analysis}

To evaluate the computational efficiency of the proposed DiffTR-Flow3D, we compare its model size (Params), inference latency, and throughput against those of other competitors on the TR-Flow3D dataset.
Compared to other coarse-to-fine algorithms (\ie, GMSF, FLOT, 3DFlow, and PointPWC-Net), our approach requires more parameters and incurs greater computational overhead; however, it achieves substantially superior performance, outperforming 3DFlow by 35.41 percentage points in ACC Strict (98.48\% vs. 63.07\%).
Furthermore, although our method possesses a larger parameter footprint than iterative refinement-based algorithms (\ie, GotFlow3D and PV-RAFT), it exhibits significantly higher efficiency during inference (\eg, achieving a latency of 52.20 ms vs. 74.01 ms and a throughput of 19.16 samples/s vs. 13.51 samples/s compared to GotFlow3D).
This stems from the fact that while PV-RAFT and GotFlow3D require 32 and 8 iterations respectively, our model requires only 2 sampling steps.

To ensure a fair and comprehensive comparison, we scaled up the parameter capacities of 3DFlow (representing coarse-to-fine baselines) and GotFlow3D (representing iterative refinement schemes) to match DiffTR-Flow3D, denoting these scaled variants as 3DFlow$^{\ast}$ and GotFlow3D$^{\ast}$, respectively.
The results demonstrate that while scaling up parameters yields performance gains for both baselines, our proposed method also maintains a significant competitive edge.

The efficiency comparisons in Table~\ref{tab:efficiency} are obtained under the TR-Flow3D evaluation setting, whereas the computational requirements for larger particle sets depend on how individual modules scale with particle number.
In particular, the dense self- and cross-attention operations in the global-correlation module (Table~\ref{tab:para_config}) require $O[C(N_s^2+N_t^2+N_sN_t)]$ computation and $O(N_s^2+N_t^2+N_sN_t)$ storage for the attention matrices, where $N_s$ and $N_t$ are the source and target particle numbers, respectively, and $C$ is the feature dimension.
The neighborhood aggregation in feature extraction and local correlation scales linearly with particle number for fixed neighborhood size and feature dimension; however, the current implementation identifies the $k$ nearest neighbors by brute force from the full pairwise distance matrix, which also requires $O(N_s^2+N_t^2)$ computation and storage.
These operations therefore impose increasing computational and memory demands when extending the method to larger particle sets in high-density 3D PTV.
For particle numbers exceeding the single-pass capacity, a practical option is to process overlapping sub-volumes, as adopted in domain-decomposition-based PTV pipelines~\cite{mercier2024robust,le2025consensus}; a systematic evaluation of this extension for DiffTR-Flow3D is left for future work.

\begin{table*}[!h]
    \centering
    \caption{\textbf{Efficiency analysis on the TR-Flow3D dataset.} Inference time and throughput are measured on a single NVIDIA RTX 4090. The superscript $\ast$ indicates a parameter-scaled variant.}
    \label{tab:efficiency}
	\small
	\setlength{\tabcolsep}{4pt}
    \begin{tabular}{l|ccc|ccc}
        \toprule
        Model & \makecell[c]{Params\\(M)} & \makecell[c]{Latency\\(ms) ($\downarrow$)} & \makecell[c]{Throughput\\(samples/s) ($\uparrow$)} & \makecell[c]{EPE$\downarrow$} & \makecell[c]{ACC Strict$\uparrow$} & \makecell[c]{ACC Relax$\uparrow$} \\
        \midrule
        GMSF          & 7.07	& 38.15  & 26.22 & 0.18789& 25.11\%& 33.03\% \\
        FLOT          & 0.11	& 13.45  & 74.34 & 0.06051& 41.03\%& 54.54\% \\
        PointPWC-Net  & 7.72	& 35.65  & 28.05 & 0.07101& 50.14\%& 63.61\% \\
        PV-RAFT       & 0.19	& 291.42 &  3.43 & 0.06379& 32.09\%& 49.41\%  \\
        3DFlow        & 4.77	& 30.37  & 32.93 & 0.06360& 63.07\%& 75.52\% \\
        GotFlow3D     & 1.44	& 74.01 &  13.51 & 0.02015& 85.26\%& 90.31\% \\ \hline
        3DFlow$^{\ast}$       & 10.64	& 60.07 & 16.65 & 0.02273 & 87.32\% & 91.64\% \\
        GotFlow3D$^{\ast}$    & 12.86	& 98.07 & 10.20 & 0.01777 & 90.57\% & 95.84\% \\ \hline
        DiffTR-Flow3D & 9.28	& 52.20  & 19.16 & \textbf{0.00316}& \textbf{98.48\%}& \textbf{98.72\%} \\
        \bottomrule
    \end{tabular}
\end{table*}

\subsection{Lagrangian statistics validation}
\begin{figure}[h]
 	\begin{center}
 	\includegraphics[width=1.0\columnwidth]{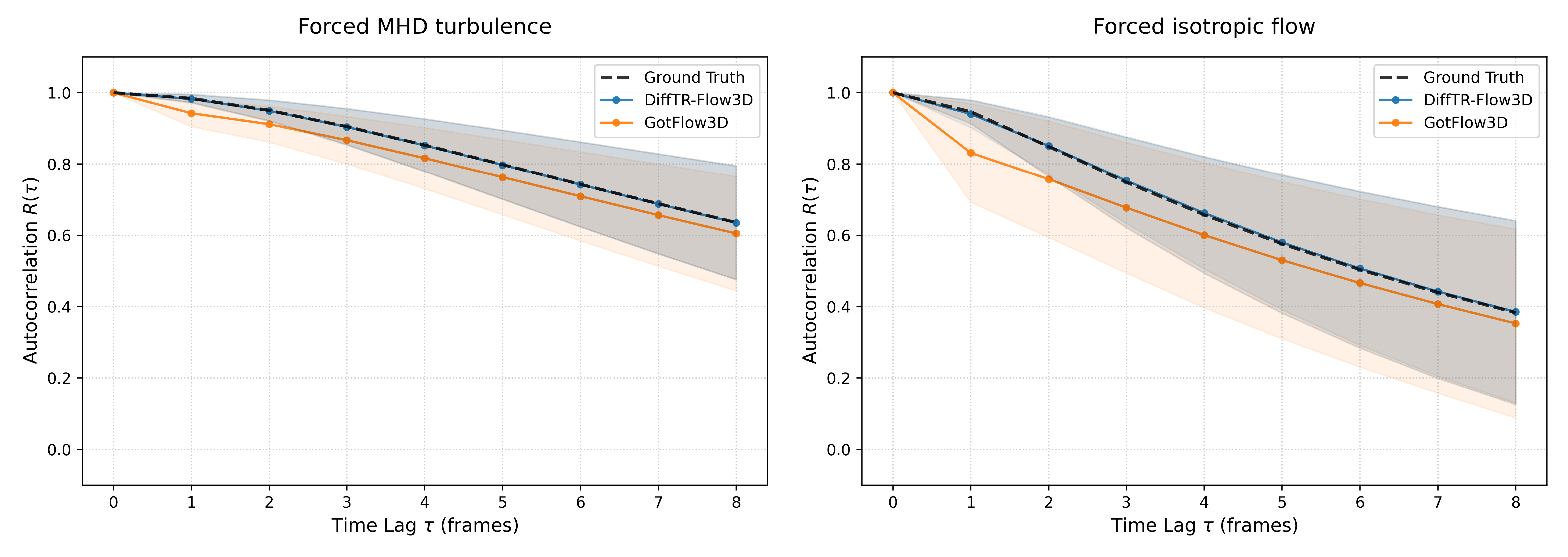}
 	\end{center}
 	\caption{\textbf{Lagrangian velocity autocorrelation with variance for forced MHD turbulence (left) and forced isotropic turbulence (right).} }
 	\label{fig:autocorrelation}
\end{figure}

To evaluate the temporal consistency and trajectory accuracy of the model, we utilize the Lagrangian Velocity Autocorrelation Function, denoted as $R(\tau)$, which measures the correlation of a particle's velocity over increasing time lags.
Since the time interval between consecutive steps is constant, we define $R(\tau)$ directly using the Lagrangian displacement $\mathbf{u}(t)$ of a particle from time $t$ to $t+1$:
\begin{equation}
R(\tau) = \frac{\langle \mathbf{u}(t) \cdot \mathbf{u}(t + \tau) \rangle}{\langle \mathbf{u}(t) \cdot \mathbf{u}(t) \rangle},
\end{equation}
where $\tau$ indicates the time lag (in frames) and $\langle \cdot \rangle$ denotes the ensemble average over all particles and initial time steps.
As shown in Fig. \ref{fig:autocorrelation}, DiffTR-Flow3D achieves a high degree of alignment with the ground truth across both forced MHD turbulence and forced isotropic turbulence, successfully capturing the mean trend and the variance distribution, whereas GotFlow3D exhibits a notable divergence from the ground truth.

\subsection{Additional results on unseen scenarios}
To further validate the efficacy of our approach, we provide additional comparisons with STB and GotFlow3D on the CylinderFlow benchmark and compare them with DiffTR-Flow3D-enhanced pipelines, as shown in Table \ref{tab:app_cylinderflow}.
While GotFlow3D achieves initial correspondence between the first and second frames comparable to ours in most settings, its performance degrades significantly in subsequent tracking steps, yielding lower overall accuracy. In contrast, our proposed method maintains robust and high accuracy throughout the entire tracking sequence.
Although STB \cite{schanz2016shake,Jahn2021} remains the gold standard for time-resolved tracking under its intended small-displacement assumptions, our results suggest that large-displacement regimes remain a formidable frontier, where our method can serve as a valuable complement, extending tracking capabilities to scenarios involving substantial inter-frame motion.

\begin{figure}[h]
 	\begin{center}
 	\includegraphics[width=1.0\columnwidth]{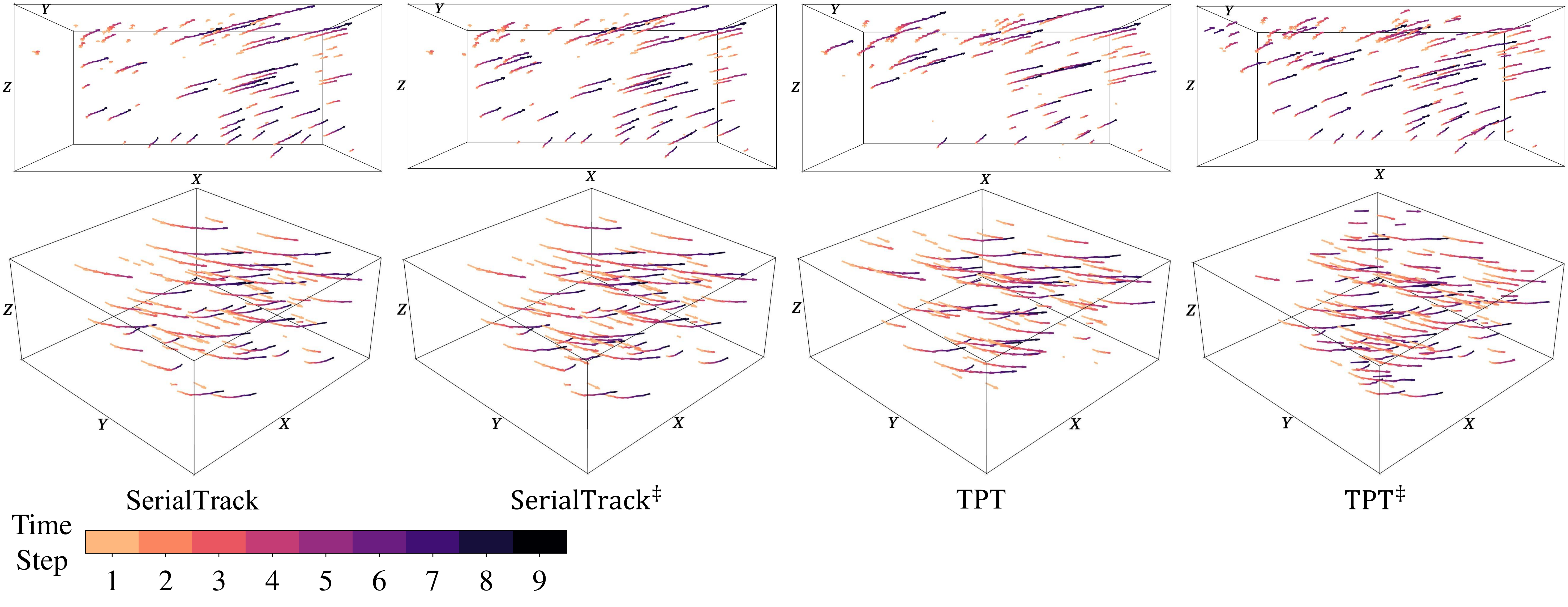}
 	\end{center}
 	\caption{\textbf{GotFlow3D applied to the mechanical shear testing performed on hydrogels to characterize their viscoelastic behavior under shear deformation.} The superscript $\ddag$ indicates the incorporation of GotFlow3D into the SerialTrack and T-PT pipelines.}
 	\label{fig:experimetal_Gotflow3D}
\end{figure}

Furthermore, in mechanical shear testing of hydrogels (Fig. \ref{fig:experimetal_Gotflow3D}), GotFlow3D exhibits limited efficacy. When integrated with T-PT, the initial tracked particle count increases (106 \vs 88 baseline);
however, the number of tracking particles decays rapidly, ending at 33 (\vs 21 baseline).
A similar trend is observed for SerialTrack (95 initial vs. 89 baseline, dropping to 32 \vs 23), confirming that despite a higher initial count, GotFlow3D struggles to maintain multi-frame tracking stability.

\begin{table}[t]
\caption{\textbf{Extended tracking accuracy $E_y$ (detailed in Eq. \ref{eq:tracking_acc}) for the CylinderFlow dataset, including STB and GotFlow3D variants.}}
\label{tab:app_cylinderflow}%
\begin{tabular}{@{}l|lll|lll|lll@{}}
\toprule %
& \multicolumn{3}{c|}{$\Delta t=30$}& \multicolumn{3}{c|}{$\Delta t=60$}& \multicolumn{3}{c}{$\Delta t=90$}\\
\multirow{-2}{*}{$n_p=3,940$}& \texttt{First}& \texttt{Last}& \texttt{Mean}& \texttt{First}& \texttt{Last}& \texttt{Mean}& \texttt{First}& \texttt{Last}& \texttt{Mean}\\
\midrule
STB& 92.91& 11.21& 55.10& 89.15& 14.60& 57.83& 82.67& 15.87& 55.87 \\
SerialTrack& 98.77& 70.69& 82.84& 94.80& 61.35& 75.18& 90.17& 52.95& 67.67 \\
$\text{SerialTrack}^{\ddag}$& 99.51& 85.74& 92.84& 98.56& 83.36& 90.95& 96.08& 76.46& 86.70 \\
$\text{SerialTrack}^{\dag}$& 99.56& 90.89& 94.59& 98.38& 89.99& 93.42& 96.89& \textbf{87.28}& 91.52 \\
T-PT& 99.72& 95.38& 97.43& 95.67& 79.49& 86.62& 82.54& 58.00& 67.67 \\
$\text{T-PT}^{\ddag}$& 99.82& 96.00& 98.11& 99.43& 92.56& 96.33& 97.82& 71.18& 86.78 \\
$\text{T-PT}^{\dag}$& \textbf{100.00}& \textbf{98.16}& \textbf{99.26}& \textbf{99.92}& \textbf{96.85}& \textbf{98.63}& \textbf{99.58}& 82.20& \textbf{93.40} \\
\midrule

& \multicolumn{3}{c|}{$\Delta t=30$}& \multicolumn{3}{c|}{$\Delta t=60$}& \multicolumn{3}{c}{$\Delta t=90$}\\
\multirow{-2}{*}{$n_p=9,850$}& \texttt{First}& \texttt{Last}& \texttt{Mean}& \texttt{First}& \texttt{Last}& \texttt{Mean}& \texttt{First}& \texttt{Last}& \texttt{Mean}\\
\midrule
STB& 86.52& 7.27& 47.46& 81.44& 9.57& 49.01& 69.67& 10.07& 43.39 \\
SerialTrack& 97.92& 68.52& 81.00& 94.69& 59.02& 73.62& 88.76& 48.93& 65.11 \\
$\text{SerialTrack}^{\ddag}$& 98.98& 75.58& 86.80& 97.30& 69.57& 83.09& 92.30& 58.87& 74.90 \\
$\text{SerialTrack}^{\dag}$& 99.45& 88.08& 93.80& 98.24& 87.19& 92.60& 95.89& \textbf{83.04}& \textbf{89.41} \\
T-PT& 99.65& 93.50& 96.48& 91.18& 70.34& 79.31& 73.36& 50.64& 59.32 \\
$\text{T-PT}^{\ddag}$& 99.62& 93.72& 96.49& 97.97& 75.60& 87.28& 87.19& 53.84& 67.60 \\
$\text{T-PT}^{\dag}$& \textbf{99.99}& \textbf{97.58}& \textbf{99.02}& \textbf{99.76}& \textbf{89.97}& \textbf{96.71}& \textbf{98.34}& 74.22& 87.38 \\
\midrule

& \multicolumn{3}{c|}{$\Delta t=30$}& \multicolumn{3}{c|}{$\Delta t=60$}& \multicolumn{3}{c}{$\Delta t=90$}\\
\multirow{-2}{*}{$n_p=19,700$}& \texttt{First}& \texttt{Last}& \texttt{Mean}& \texttt{First}& \texttt{Last}& \texttt{Mean}& \texttt{First}& \texttt{Last}& \texttt{Mean}\\
\midrule
STB& 77.38& 5.21& 39.73& 64.94& 6.80& 36.70& 48.41& 6.38& 28.23 \\
SerialTrack& 97.97& 69.08& 81.98& 93.99& 59.38& 73.92& 88.15& 50.13& 65.48 \\
$\text{SerialTrack}^{\ddag}$& 98.37& 69.44& 82.65& 95.18& 60.49& 75.53& 89.81& 49.77& 67.09 \\
$\text{SerialTrack}^{\dag}$& 99.24& 85.75& 92.95& 97.84& \textbf{82.81}& 90.28& \textbf{93.77}& \textbf{73.96}& \textbf{83.58} \\
T-PT& 98.96& 91.86& 95.24& 86.61& 65.74& 74.34& 67.96& 50.08& 56.98 \\
$\text{T-PT}^{\ddag}$& 99.40& 91.69& 95.25& 92.41& 65.74& 76.33& 73.19& 50.08& 57.83 \\
$\text{T-PT}^{\dag}$& \textbf{99.93}& \textbf{96.94}& \textbf{98.74}& \textbf{99.35}& 80.11& \textbf{91.48}& 90.86& 62.78& 76.03 \\
\botrule
\end{tabular}
The superscripts $\dag$ and $\ddag$ indicate the incorporation of DiffTR-Flow3D and GotFlow3D, respectively, into the SerialTrack and T-PT pipelines.
The best results are highlighted in bold; $\Delta t$ signifies the time interval between frames; $n_p$ indicates the number of particles sampled from the CylinderFlow dataset. \texttt{First} denotes the tracking accuracy of the initial trajectory segment (between the 1st and 2nd frames); \texttt{Last} measures the association accuracy between the first and final frames; and \texttt{Mean} represents the average tracking accuracy computed across all time steps.
\end{table}

\subsection{Comparative analysis across various flow field patterns}
We provide detailed performance comparisons against deep learning-based competitors across various flow types.
As demonstrated in Table \ref{tab:comparison_pattern}, DiffTR-Flow3D outperforms all baseline models across all evaluation metrics and flow field patterns.
Notably, while GotFlow3D remains competitive in simpler scenarios, a significant performance gap emerges in more complex fields such as Beltrami, Forced MHD, and Forced Isotropic flows.
This highlights the overall superiority of our proposed architecture in handling challenging scenarios.

\begin{table}[h]
\centering
\caption{\textbf{Comparison of performance metrics across various flow field patterns on the TR-Flow3D testing dataset.}}
\label{tab:comparison_pattern}
\footnotesize
\setlength{\tabcolsep}{3pt}
\begin{tabular}{@{}llcccc@{}}
\toprule
\textbf{Model} & \textbf{Flow Field Pattern} & \textbf{EPE} $\downarrow$ & \textbf{ACC Strict} $\uparrow$ & \textbf{ACC Relax} $\uparrow$ & \textbf{Outliers} $\downarrow$ \\ \midrule

\multirow{6}{*}{GMSF} & Uniform flow & 0.02591 & 87.17\% & 89.54\% & 10.46\% \\
 & Transitional boundary flow & 0.03255 & 75.93\% & 84.91\% & 15.09\% \\
 & Turbulent channel flow  & 0.07480 & 19.23\% & 52.14\% & 47.86\% \\
 & Beltrami flow  & 0.39621 & 0.09\% & 0.49\% & 99.51\% \\
 & Forced MHD turbulence  & 0.22258 & 0.08\% & 0.28\% & 99.72\% \\
 & Forced isotropic turbulence & 0.28831 & 0.11\% & 0.58\% & 99.42\% \\ \midrule

\multirow{6}{*}{FLOT} & Uniform flow & 0.00842 & 91.99\% & 99.32\% & 0.68\% \\
 & Transitional boundary flow & 0.01087 & 90.55\% & 97.86\% & 2.14\% \\
 & Turbulent channel flow  & 0.02565 & 57.14\% & 88.86\% & 11.14\% \\
 & Beltrami flow & 0.11265 & 31.17\% & 55.97\% & 44.03\% \\
 & Forced MHD turbulence & 0.07652 & 0.35\% & 2.33\% & 97.67\% \\
 & Forced isotropic turbulence & 0.10256 & 0.88\% & 5.48\% & 94.52\% \\ \midrule

\multirow{6}{*}{PV-RAFT} & Uniform flow & 0.01054 & 84.68\% & 96.70\% & 3.30\% \\
 & Transitional boundary flow & 0.01238 & 87.25\% & 96.08\% & 3.92\% \\
 & Turbulent channel flow & 0.03207 & 45.89\% & 80.80\% & 19.20\% \\
 & Beltrami flow & 0.12118 & 18.15\% & 50.41\% & 49.59\% \\
 & Forced MHD turbulence & 0.07409 & 0.46\% & 3.05\% & 96.95\% \\
 & Forced isotropic turbulence & 0.09885 & 1.28\% & 7.44\% & 92.56\% \\ \midrule

\multirow{6}{*}{3DFlow} & Uniform flow & 0.14318 & 17.09\% & 33.88\% & 66.12\% \\
 & Transitional boundary flow & 0.01688 & 85.00\% & 96.10\% & 3.90\% \\
 & Turbulent channel flow & 0.01010 & 90.53\% & 98.33\% & 1.67\% \\
 & Beltrami flow & 0.13660 & 53.62\% & 63.86\% & 36.14\% \\
 & Forced MHD turbulence & 0.01147 & 63.73\% & 81.32\% & 18.68\% \\
 & Forced isotropic turbulence & 0.09688 & 49.70\% & 63.24\% & 36.76\% \\ \midrule

\multirow{6}{*}{\shortstack[l]{PointPWC-\\Net}} & Uniform flow & 0.07532 & 27.76\% & 46.67\% & 53.33\% \\
 & Transitional boundary flow & 0.01837 & 87.01\% & 93.13\% & 6.87\% \\
 & Turbulent channel flow & 0.00862 & 94.27\% & 98.30\% & 1.70\% \\
 & Beltrami flow & 0.13098 & 46.40\% & 59.85\% & 40.15\% \\
 & Forced MHD turbulence & 0.03482 & 27.91\% & 54.95\% & 45.05\% \\
 & Forced isotropic turbulence & 0.15539 & 10.47\% & 23.68\% & 76.32\% \\ \midrule

\multirow{6}{*}{GotFlow3D} & Uniform flow & 0.00422 & 98.49\% & 99.90\% & 0.10\% \\
 & Transitional boundary flow & 0.00373 & 99.67\% & 99.96\% & 0.04\% \\
 & Turbulent channel flow & 0.00427 & 98.57\% & 99.56\% & 0.44\% \\
 & Beltrami flow & 0.05449 & 76.44\% & 84.56\% & 15.44\% \\
 & Forced MHD turbulence & 0.01127 & 73.81\% & 85.52\% & 14.48\% \\
 & Forced isotropic turbulence & 0.03246 & 73.85\% & 79.88\% & 20.12\% \\ \midrule

\multirow{6}{*}{\shortstack[l]{DiffTR-\\Flow3D}} & Uniform flow & 0.00028& $100.0$\%& $100.0$\%&  $0.00$\% \\
 & Transitional boundary flow & 0.00078& 99.99\%&  $100.0$\%&  $0.00$\% \\
 & Turbulent channel flow & 0.00078& 99.91\%& 99.97\%& 0.03\% \\
 & Beltrami flow & 0.00390& 98.43\%& 98.79\%& 1.21\% \\
 & Forced MHD turbulence & 0.00148& 97.55\%& 98.66\%& 1.34\% \\
 & Forced isotropic turbulence & 0.00803& 95.68\%& 96.24\%& 3.76\% \\ \botrule
\end{tabular}
\end{table}

\subsection{Kinematic consistency of the hydrogel tracking results}
\label{sec:supp-hydrogel-consistency}
The hydrogel was sheared quasi-statically in nominally equal increments of 4\% engineering shear strain~\cite{yang2022serialtrack}.
Over the small imaged region, the motion between consecutive frames should therefore be well approximated by a single affine map,
\begin{equation}
\mathbf{x}_{k+1}\approx\mathbf{F}_k\mathbf{x}_k+\mathbf{t}_k ,
\end{equation}
where $\mathbf{F}_k$ is an effective incremental deformation gradient and $\mathbf{t}_k$ is a rigid translation.
This homogeneous model provides an approximate description of the local motion. An incorrect association may produce a displacement that is inconsistent with this model, resulting in a large residual.

We fitted the model separately for each of the nine frame intervals by ordinary least squares.
The fit used only those associations that are identical across all four raw outputs, namely SerialTrack~\cite{yang2022serialtrack}, SerialTrack$^\dagger$, T-PT~\cite{patel2018rapid} and T-PT$^\dagger$, which yields 61--82 associations per interval.
No residual trimming was applied in this analysis.
For each association retained in Fig.~6 of the main text, we then computed the residual
\begin{equation}
r=\lVert\mathbf{x}_{k+1}-\mathbf{F}_k\mathbf{x}_k-\mathbf{t}_k\rVert .
\end{equation}
A correct association should yield a small residual, whereas an incorrect association, which links the particle to a different one, should yield a large residual.
The fitted model provides a common reference for assessing whether additional associations follow the predominant motion.
Because it is estimated from tracking outputs, this analysis evaluates kinematic consistency rather than independently establishing association accuracy.

\begin{table}[h]
\centering\small
\caption{Residuals $r$ (voxels) from the per-interval affine model, pooled over nine frame intervals for trajectories continuously connected to the first frame (Fig.~6).
For each comparison, \emph{Shared} denotes associations present in both pipelines, \emph{Added} those exclusive to the DiffTR-integrated pipeline, and \emph{Removed} those exclusive to the baseline.}
\label{tab:r1-hydrogel}
\begin{tabular}{llrrrr}
\toprule
Comparison & Associations & $n$ & Median & P90 & Max \\
\midrule
SerialTrack / SerialTrack$^\dagger$ & Shared  & 440 & 1.88 & 3.36 & 4.35 \\
                                    & Added   & 201 & 1.59 & 3.19 & 4.46 \\
                                    & Removed &  29 & 1.57 & 3.70 & 5.46 \\
\midrule
T-PT / T-PT$^\dagger$               & Shared  & 431 & 1.75 & 3.22 & 4.46 \\
                                    & Added   & 325 & 1.91 & 3.43 & 5.46 \\
                                    & Removed &  11 & 1.43 & 2.61 & 4.91 \\
\bottomrule
\end{tabular}
\end{table}

Table~\ref{tab:r1-hydrogel} compares the associations that are added or removed when DiffTR-Flow3D is integrated with those shared by both pipelines.
Each retained trajectory contributes one association per frame interval it spans, so n counts frame-to-frame links pooled over the nine intervals rather than particles.
The added associations have median residuals of 1.59 voxels for SerialTrack and 1.91 voxels for T-PT, comparable to 1.88 and 1.75 voxels for the shared associations, and their maximum residuals (4.46 and 5.46 voxels) are of the same order.
All residuals remain far below the nearest-neighbor distance between detected particles (median 95.80 voxels; minimum 15.55 voxels), which sets the scale of the residual expected from linking to a different particle.
The integration removes only a few baseline associations (29 and 11), compared with the 201 and 325 associations it adds.

These results indicate that the additional associations follow the predominant motion as closely as the shared associations.
This suggests that the increase in retained trajectories in Fig.~6 largely reflects plausible associations, and thus provides a reasonable indication of the improvement offered by DiffTR-Flow3D in this experiment.

\clearpage
\section{More implementation details}\label{secB}

\setcounter{figure}{0}
\setcounter{table}{0}
\setcounter{algorithm}{0}

\subsection{Pseudocode of DiffTR-Flow3D}
Algorithm \ref{alg:training} outlines the comprehensive training procedure for DiffTR-Flow3D.
The process begins with a diffusion forward step, where a noisy flow trajectory $D_n$ is generated to perturb the source particles.
The core architecture then follows a coarse-to-fine estimation strategy: first, a coarse motion approximation is derived from global context features; second, a fine-grained refinement is performed by warping the source points and enhancing features via the Local-Global Attention.
Finally, the model is optimized using a robust regression loss and the AdamW optimizer under a OneCycle learning rate schedule to ensure stable convergence.

\begin{algorithm}[h]
\caption{Training Procedure of DiffTR-Flow3D}
\label{alg:training}
\begin{algorithmic}[1]
\Require Training dataset $\mathcal{D}$, Max epochs $E$, Learning rate $\eta$, Batch size $B$, Diffusion steps $N$.
\Ensure Optimized model parameters $\theta^*$.

\State Initialize model parameters $\theta$ (including $\phi_e, \phi_r, W_q, W_k, \phi_\mathcal{T}$, etc.).

\For{epoch $e = 1$ to $E$}
    \For{batch $\{(P_s, P_t, D_{gt})\}$ sampled from $\mathcal{D}$}
        \State \textbf{// Diffusion Forward Process}
        \State Sample timestep $n \sim \text{Uniform}(1, N)$ and noise $\boldsymbol{\epsilon} \sim \mathcal{N}(0, I)$.
        \State Compute noisy flow trajectory $D_{n}$ using $D_{gt}, \boldsymbol{\epsilon}, \bar{\alpha}_n$. \Comment{Eq. \ref{eq:diff}}
        \State Perturb source particles: $P'_s \leftarrow P_s + D_n$.

        \State \textbf{// Step 1: Coarse Motion Estimation}
        \State Extract context features $F_s^e, F_t^e$ from $\{P'_s, P_t\}$. \Comment{Eq. \ref{eq:7}}
        \State Compute coarse correlation matrix $D_c^e$. \Comment{Eq. \ref{eq:8}}
        \State Estimate coarse flow approximation $D_{coarse}$. \Comment{Eq. \ref{eq:9}}

        \State \textbf{// Step 2: Fine Motion Refinement}
        \State Pre-align source points $\hat{P}_s \leftarrow P'_s + D_{coarse}$.
        \State Extract refined features $F_s^r, F_t^r$ from $\{\hat{P}_s, P_t\}$.
        \State Enhance features via Local-Global Attention. \Comment{Eq. \ref{eq:10}, \ref{eq:11}}
        \State Predict residual flow $D_{pred}$. \Comment{Eq. \ref{eq:8}, \ref{eq:9}}
        \State Compose clean flow estimate $\hat{D}_{s \rightarrow t} \leftarrow D_n + D_{coarse} + D_{pred}$.

        \State \textbf{// Optimization}
        \State Compute the robust regression loss (with $\delta=0.01$):
        \State \quad $\mathcal{L} = \sum_{i}(\lVert \hat{D}_{s \rightarrow t}(i)-D_{gt}(i)\rVert_{1} + \delta)^{q}$ \Comment{Eq. \ref{eq:12}}
        \State Compute gradients: $g \leftarrow \nabla_\theta \mathcal{L}$.
        \State Update parameters: $\theta \leftarrow \text{AdamW}(\theta, g, \eta)$.
    \EndFor
    \State Update learning rate via OneCycle schedule.
\EndFor
\State \textbf{return} $\theta^*$
\end{algorithmic}
\end{algorithm}

\subsection{Parameter configuration for DiffTR-Flow3D}
The specific architectural parameters and layer configurations for DiffTR-Flow3D are summarized in Table \ref{tab:para_config}.
We employ LeakyReLU for the initial feature extraction layers and transition to ReLU and GELU activations for the correlation modules.
Notably, the Global Correlation module is the most computationally dense part of the model, featuring an 8-layer attention-based structure to aggregate spatial information.

\begin{table}[h]
\centering
\caption{\textbf{Parameter configuration for DiffTR-Flow3D.}}
\small
\renewcommand{\arraystretch}{1.0} %
\begin{tabular}{lcccc}
\toprule
\textbf{\#} & \textbf{Operation} & \textbf{Feature Size} & \textbf{Kernel/Dim.} & \textbf{Activation} \\
\midrule
\multicolumn{5}{l}{\textit{Feature Extraction (0.08M)} $\times 2$} \\
1 & Conv2d & $64 \times N \times 16$ & $1 \times 1$ & LeakyReLU \\
2 & MaxPool & $64 \times N$ & - & - \\
3 & Conv2d & $64 \times N \times 16$ & $1 \times 1$ & LeakyReLU \\
4 & MaxPool & $64 \times N$ & - & - \\
5 & Conv2d & $64 \times N \times 16$ & $1 \times 1$ & LeakyReLU \\
6 & MaxPool & $64 \times N$ & - & - \\
7 & Conv2d & $128 \times N \times 16$ & $1 \times 1$ & LeakyReLU \\
8 & MaxPool & $128 \times N$ & - & - \\
9 & Concat & $320 \times N$ & - & - \\
10 & Conv1d & $128 \times N$ & 1 & LeakyReLU \\
\midrule
\multicolumn{5}{l}{\textit{Motion Estimation (0.03M)} $\times 2$} \\
1 & Linear & $128 \times N$ & $128 \to 128$ & - \\
2 & Linear & $128 \times N$ & $128 \to 128$ & - \\
\midrule
\multicolumn{5}{l}{\textit{Local Correlation (1.71M)} $\times 1$} \\
1 & Linear & $N \times 128$ & $128 \to 128$ & - \\
2 & Linear & $N \times 128$ & $128 \to 128$ & - \\
3 & Linear & $N \times K \times 128$ & $128 \to 128$ & - \\
4 & MLP & $N \times K \times 128$ & $3 \to 128$ & ReLU \\
5 & MLP & $N \times K \times 128$ & $128 \to 128$ & ReLU \\
6 & Linear & $N \times 128$ & $128 \to 128$ & - \\
\midrule
\multicolumn{5}{l}{\textit{Global Correlation (7.35M)}} \\
\multicolumn{5}{l}{\texttt{self-attention} $\times 8$} \\
1 & Linear & $N \times 128$ & $128 \to 128$ & - \\
2 & Linear & $N \times 128$ & $128 \to 128$ & - \\
3 & Linear & $N \times K \times 128$ & $128 \to 128$ & - \\
\multicolumn{5}{l}{\texttt{cross-attention} $\times 8$} \\
1 & Linear & $N \times 128$ & $128 \to 128$ & - \\
2 & Linear & $N \times 128$ & $128 \to 128$ & - \\
3 & Linear & $N \times K \times 128$ & $128 \to 128$ & - \\
4 & MLP & $N \times K \times 128$ & $128 \to 128$ & GELU \\
\bottomrule
\end{tabular}
\label{tab:para_config}
\end{table}

\clearpage
\section{More visualizations}\label{secC}

\subsection{Visualizations on the TR-Flow3D dataset}
\begin{figure}[h]
 	\begin{center}
 	\includegraphics[width=1.0\columnwidth]{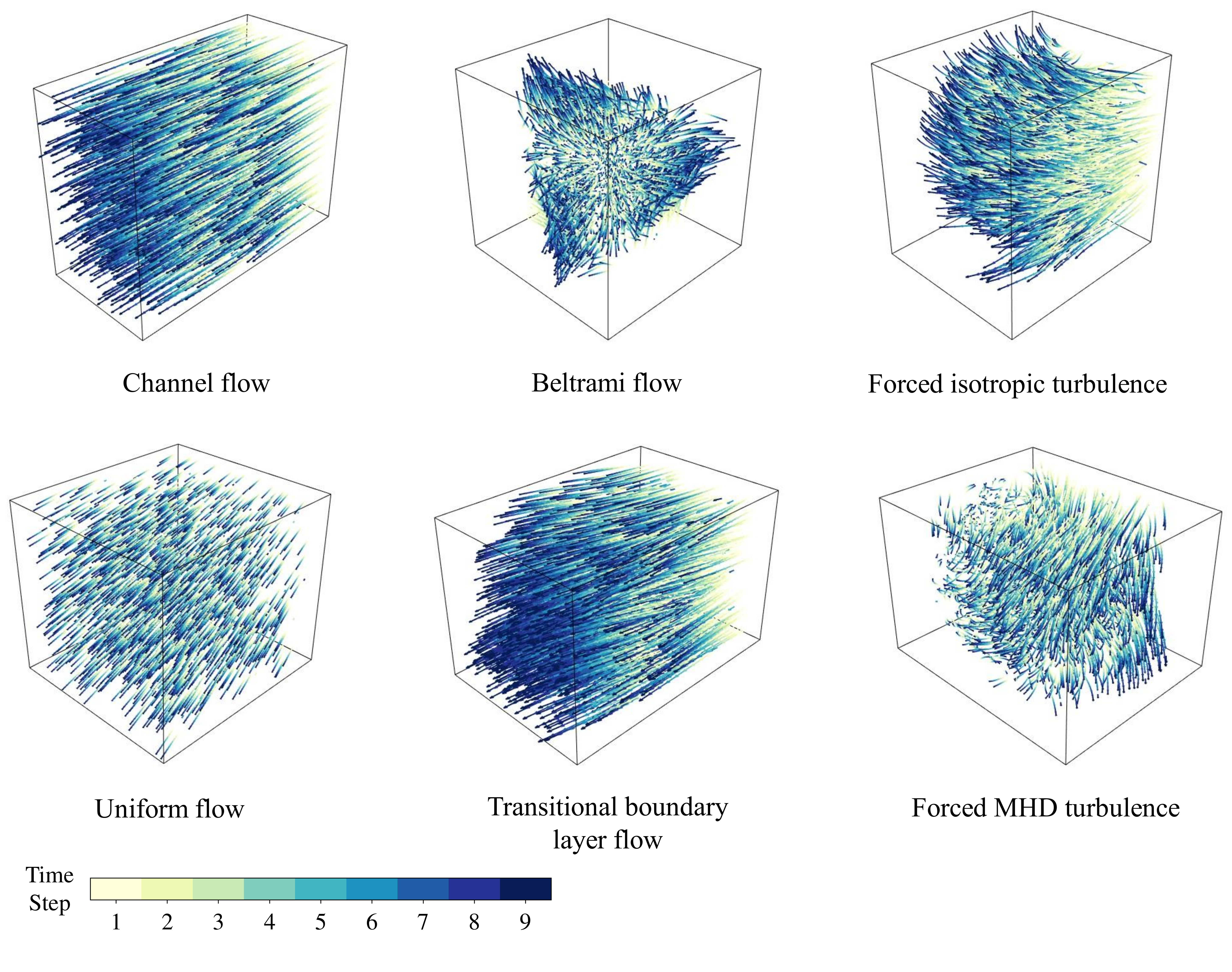}
 	\end{center}
 	\caption{\textbf{Examples of the six flow patterns on the TR-Flow3D dataset.}}
 	\label{fig:more_visualization}
\end{figure}

We provide additional qualitative visualizations to further illustrate the diversity and complexity of the TR-Flow3D dataset.
As shown in Fig. \ref{fig:more_visualization}, we present six representative flow patterns that characterize various fluid dynamic behaviors.

\end{appendices}

\end{document}